\documentclass[twocolumn]{aastex63}

\usepackage{amsmath}
\usepackage{amssymb}
\usepackage{euscript}
\usepackage{algorithm2e,algorithmic}
\usepackage{bbold}
\usepackage{paralist}
\usepackage{color}
\usepackage{amsmath}
\usepackage{lineno}
\usepackage{comment}
\usepackage{soul}

\submitjournal{ApJ}

\shorttitle{SN\,2018\,fif}
\shortauthors{Soumagnac et al.}

\begin{document}


\title{SN\,2018\,fif: the explosion of a large red supergiant discovered in its infancy by the Zwicky Transient Facility}

\correspondingauthor{Maayane T. Soumagnac}
\email{mtsoumagnac@lbl.gov}
\author[0000-0001-6753-1488]{Maayane T. Soumagnac}
\affiliation{Department of Particle Physics and Astrophysics, Weizmann Institute of Science, Rehovot 76100, Israel}
\affiliation{Lawrence Berkeley National Laboratory, 1 Cyclotron Road, Berkeley, CA 94720, USA}
\author{Noam Ganot}
\affiliation{Department of Particle Physics and Astrophysics, Weizmann Institute of Science, Rehovot 76100, Israel}
\author{Ido Irani}
\affiliation{Department of Particle Physics and Astrophysics, Weizmann Institute of Science, Rehovot 76100, Israel}
\author{Avishay Gal-yam}
\affiliation{Department of Particle Physics and Astrophysics, Weizmann Institute of Science, Rehovot 76100, Israel}
\author{Eran O. Ofek}
\affiliation{Department of Particle Physics and Astrophysics, Weizmann Institute of Science, Rehovot 76100, Israel}
\author{Eli Waxman}
\affiliation{Department of Particle Physics and Astrophysics, Weizmann Institute of Science, Rehovot 76100, Israel}
\author{Jonathan Morag}
\affiliation{Department of Particle Physics and Astrophysics, Weizmann Institute of Science, Rehovot 76100, Israel}
\author{Ofer Yaron}
\affiliation{Department of Particle Physics and Astrophysics, Weizmann Institute of Science, Rehovot 76100, Israel}
\author{Steve Schulze}
\affiliation{Department of Particle Physics and Astrophysics, Weizmann Institute of Science, Rehovot 76100, Israel}
\author{Yi Yang}
\affiliation{Department of Particle Physics and Astrophysics, Weizmann Institute of Science, Rehovot 76100, Israel}
\author{Adam Rubin}
\affiliation{European Southern Observatory, Karl-Schwarzschild-Str 2, 85748 Garching bei Munchen, Germany}
\author{S. Bradley Cenko}
\affiliation{Astrophysics Science Division, NASA Goddard Space Flight Center, MC 661, Greenbelt, MD 20771, USA}
\affiliation{Joint Space-Science Institute, University of Maryland, College Park, MD 20742, USA}
\author{Jesper Sollerman}
\affiliation{The Oskar Klein Centre, Department of Astronomy, Stockholm University, AlbaNova, 10691 Stockholm, Sweden}
\author{Daniel A. Perley}
\affiliation{Astrophysics Research Institute, Liverpool John Moores University, 146 Brownlow Hill, Liverpool L3 5RF, UK}
\author{Christoffer Fremling}
\affiliation{Division of Physics, Mathematics and Astronomy, California Institute of Technology, Pasadena, CA 91125, USA}
\author{Peter Nugent}
\affiliation{Lawrence Berkeley National Laboratory, 1 Cyclotron Road, Berkeley, CA 94720, USA}
\affiliation{Department of Astronomy, University of California, Berkeley, CA 94720-3411, USA}
\author{James D. Neill}
\affiliation{Observatories of the Carnegie Institution for Science, 813 Santa Barbara Street, Pasadena, CA 91101, USA}
\author{Emir Karamehmetoglu}
\affiliation{The Oskar Klein Centre, Department of Astronomy, Stockholm University, AlbaNova, 10691 Stockholm, Sweden}
\author[0000-0001-8018-5348]{Eric C. Bellm}
\affiliation{DIRAC Institute, Department of Astronomy, University of Washington, 3910 15th Avenue NE, Seattle, WA 98195, USA}
\author[0000-0001-8208-2473]{Rachel J. Bruch}
\affiliation{Department of Particle Physics and Astrophysics, Weizmann Institute of Science, Rehovot 76100, Israel}
\author{Rick Burruss}
\affiliation{Caltech Optical Observatories Palomar Mountain CA 92060}
\author{Virginia Cunningham}
\affiliation{Astronomy Department, University of Maryland, College Park, MD 20742, USA}
\author{Richard Dekany}
\affiliation{Caltech Optical Observatories California Institute of Technology, Pasadena, CA  91125}
\author[0000-0001-8205-2506]{V. Zach Golkhou}%
\affiliation{DIRAC Institute, Department of Astronomy, University of Washington, 3910 15th Avenue NE, Seattle, WA 98195, USA}
\affiliation{The eScience Institute, University of Washington, Seattle, WA 98195, USA}
\author{Matthew J. Graham}
\affiliation{Division of Physics, Mathematics and Astronomy, California Institute of Technology, Pasadena, CA 91125, USA}
\author{Mansi M. Kasliwal}
\affiliation{Division of Physics, Mathematics and Astronomy, California Institute of Technology, Pasadena, CA 91125, USA}
\author{Nicholas P Konidaris}
\affiliation{Observatories of the Carnegie Institution for Science, 813 Santa Barbara Street, Pasadena, CA 91101, USA}
\author{Shrinivas R. Kulkarni}
\affiliation{Division of Physics, Mathematics and Astronomy, California Institute of Technology, Pasadena, CA 91125, USA}
\author{Thomas Kupfer}
\affiliation{Kavli Institute for Theoretical Physics, University of California, Santa Barbara, CA 93106, USA}
\author[0000-0003-2451-5482]{Russ R. Laher} %
\affiliation{IPAC, California Institute of Technology, 1200 E. California Blvd, Pasadena, CA 91125, USA}
\author[0000-0002-8532-9395]{Frank J. Masci}%
\affiliation{IPAC, California Institute of Technology, 1200 E. California Blvd, Pasadena, CA 91125, USA}
\author{Reed Riddle}
\affiliation{Caltech Optical Observatories California Institute of Technology, Pasadena, CA  91125}
\author{Mickael Rigault}
\affiliation{Universite Clermont Auvergne, CNRS/IN2P3, Laboratoire de Physique de Clermont, F-63000 Clermont-Ferrand, France}
\author[0000-0001-7648-4142]{Ben Rusholme}%
\affiliation{IPAC, California Institute of Technology, 1200 E. California Blvd, Pasadena, CA 91125, USA}
\author{Jan van Roestel}
\affiliation{Division of Physics, Mathematics and Astronomy, California Institute of Technology, Pasadena, CA 91125, USA}
\author{Barak Zackay}
\affiliation{Institute for Advanced Study, 1 Einstein Drive, Princeton, New Jersey 08540, USA}

\begin{abstract}
High cadence transient surveys are able to capture supernovae closer to their first light than before. Applying analytical models to such early emission, we can constrain the progenitor stars properties. In this paper, we present observations of SN~2018\,fif (ZTF18abokyfk). The supernova was discovered close to first light and monitored by the Zwicky Transient Facility (ZTF) and the Neil Gehrels Swift Observatory. Early spectroscopic observations suggest that the progenitor of SN~2018\,fif was surrounded by relatively small amounts of circumstellar material (CSM) compared to all previous cases. This particularity, coupled with the high cadence multiple-band coverage, makes it a good candidate to investigate using shock-cooling models. We employ the {\tt SOPRANOS} code, an implementation of the model by Sapir \& Waxman and its extension to early times by Morag, Sapir \& Waxman. Compared with previous implementations, {\tt SOPRANOS} has the advantage of including a careful account of the limited temporal validity domain of the shock-cooling model as well as allowing usage of the entirety of the early UV data. We find that the progenitor of SN~2018\,fif was a large red supergiant, with a radius of $R=744.0_{-128.0}^{+183.0}\,R_{\odot}$ and an ejected mass of $M_{\rm ej}=9.3_{-5.8}^{+0.4}\,M_{\odot}$. Our model also gives information on the explosion epoch, the progenitor inner structure, the shock velocity and the extinction. The distribution of radii is double-peaked, with lower radii corresponding to lower values of the extinction, earlier recombination times and better match to the early UV data. If these correlations persist in future objects, denser spectroscopic monitoring constraining the time of recombination, as well as accurate UV observations (e.g. with ULTRASAT), will help break the radius-extinction degeneracy and independently determine both.


\end{abstract}


\section{Introduction}
In recent years, advances in the field of high-cadence transient surveys have made it possible to systematically discover and follow-up supernovae (SNe) within hours of their first light \cite[e.g.,][]{Nugent2011,Gal-Yam2014,Yaron2017,Arcavi2017,Tartaglia2017}. This offers new opportunities to understand the early stages of core collapse (CC) SN explosions and to identify the nature of their progenitor stars.

First, rapid spectroscopic follow-up in the hours following first light has led to the detection of ``flash ionized'' emission from infant SNe \citep{Gal-Yam2014,Shivvers2015,Khazov2016,Yaron2017,Hosseinzadeh2018}. These events show prominent, transient, high-ionization recombination emission lines in their spectra, a signature of circumstellar material (CSM) ionized by the SN shock-breakout flash ("flash spectroscopy"). \cite{Khazov2016} showed that $\sim20\%$ of the SNe discovered by the Palomar Transient Factory (PTF) within 10 days of explosion are ``flashers'', while recent results from ZTF (Bruch et al, in preparation) suggest that the fraction of such events may be even higher for events observed earlier, and that CSM around CC SNe progenitors is common.

Second, observational access to the first hours following the explosion has offered a new opportunity to test theoretical models of early emission from CC SNe and constrain their progenitor properties. The handful of cases where direct pre-explosion observations of progenitors exist \citep[e.g.,][and references therein]{Smartt2015} suggest that many Type II SNe arise from red supergiants, a population of stars with radii ranging from about $100$\,R$_{\odot}$ to $1500$\,R$_{\odot}$ \citep[e.g.,][and references therein]{Levesque2017}. In recent years, theorists have developed analytical models linking SN early multi-color light curves to progenitor properties, such as radius, mass, or inner structure. Papers by \cite{Morozova2016} and \cite{Rubin2017} review and compare these models. In this paper, we use the analytical model by \cite{SW2017} (SW17) and its extension by Morag et al. 2020 (in preparation) (M20), which has two advantages. First, it accounts for bound-free absorption in the calculation of the color temperature, a feature that may have a large impact on the estimation of the progenitor radius. Second, it extends the previous results by \cite{Rabinak2011} to later times, making additional observations useful in this analysis. M20 further extends the results of SW17 to the times immediately following breakout, allowing to use all the available early data.

 \cite{Rubin2017} optimize the number of observations included in the fit based on the limited temporal validity domain of these models, but their observations were limited to the r band. To our knowledge, SN 2013fs (Yaron et al. 2017) is the only published object for which high cadence multiple-band observations are available and which was modeled using the methodology of Rubin \& Gal-Yam (2017; see section 4.3 for a detailed discussion on this aspect of the modeling).

Comparison between early observations of CC SNe and theoretical predictions from analytical models were reported previously (e.g. by \citealt{Gall2015,Gonzalez-Gaitan2015,Rubin2017, Hosseinzadeh2019}) but these authors fit only a fixed range of times in the early light curve. \cite{Rubin2017} optimize the number of observations included in the fit based on the limited temporal validity domain of these analytical models, but their observations were limited to $r$-band observations. Recently, \cite{Ricks2019,Bersten2019,Goldberg2019,Dessart2019} and \cite{Eldridge2019} have compared observations to hydrodynamic models on simulated progenitors exploded ``by hand''.

To our knowledge, SN~2013\,fs \citep{Yaron2017} is the only published object for which high cadence multiple-band observations are available and which was modeled with an analytical model using the methodology of Rubin \& Gal-Yam (2017; see section 4.3 for a detailed discussion on this aspect of the modeling). However, the spectroscopic observations of SN~2013\,fs - the best observed ``flasher'' to date - show evidence for $\sim10^{-3}\,M_{\odot}$ of confined CSM surrounding the progenitor. The presence of CSM casts doubt upon the validity of the SW17 model in this case, and perhaps could have pushed the best-fit model radius found for this object ($R= 100 - 350  R_{\odot}$) towards the lower end of the RSG radius distribution. A ``cleaner'' supernova, with no prominent signatures of CSM around the progenitor, may be a more appropriate test case for the SW17 model.  

In this paper, we present and analyse the UV and visible-light observations of SN~2018\,fif (ZTF18abokyfk), a SN first detected shortly after explosion by the Zwicky Transient Facility (ZTF; e.g., \citealt{Bellm2019,Graham2019}) as part of the ZTF extragalactic high-cadence experiment \citep{Gal-Yam2019}. 

We present the aforementioned observations of SN~2018\,fif in \S 2. In \S 3, we present our analysis of these observations, and the spectroscopic evidence making SN~2018\,fif a good candidate for modeling. \S 4 is dedicated to the modeling of the shock-cooling phase of SN~2018\,fif and the derivation of the progenitor parameters. We then summarize our main results in \S 5. 

\section{Observations and data reduction}
In this section, we present the observations of SN~2018\,fif by ZTF and the Neil Gehrels Swift Observatory (\textit{Swift}). 

\subsection{Discovery}\label{sec:discovery}

\begin{figure*}
\begin{center}
\includegraphics[scale=.35]{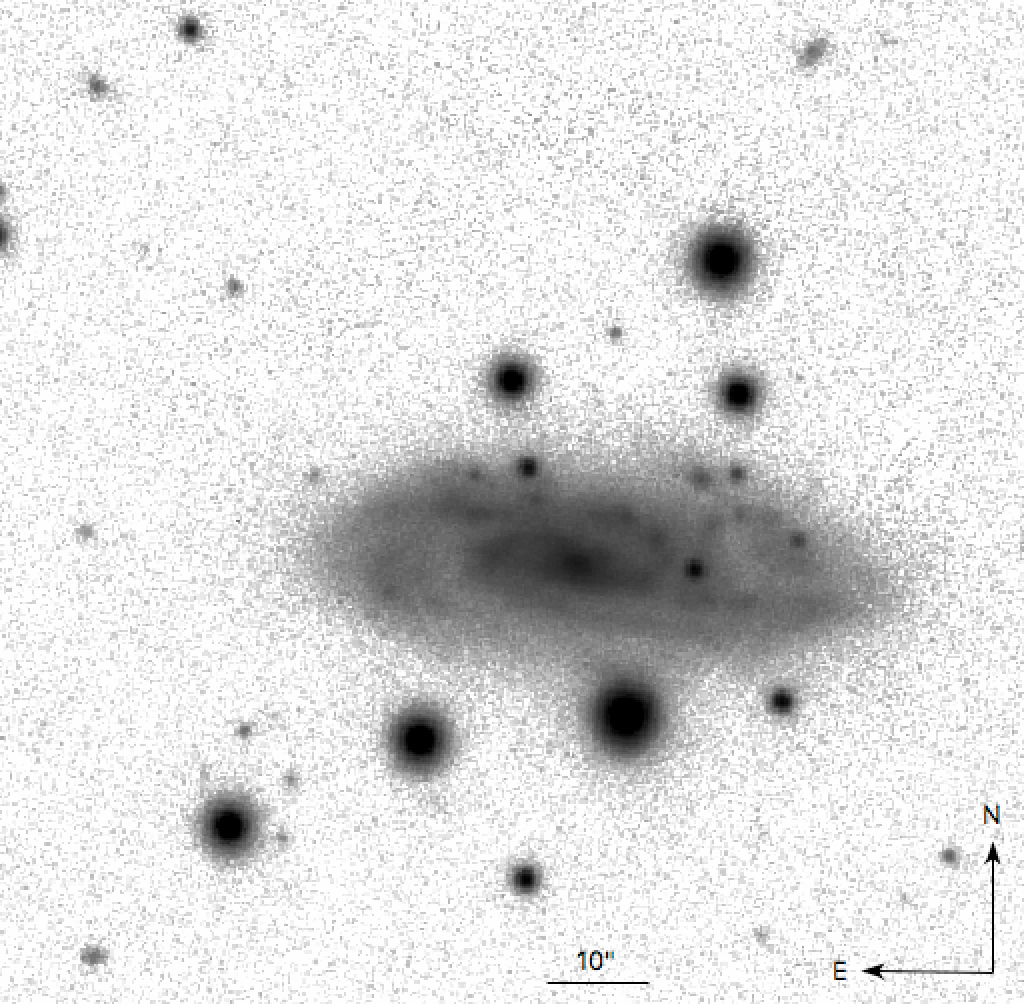}
\includegraphics[scale=.35]{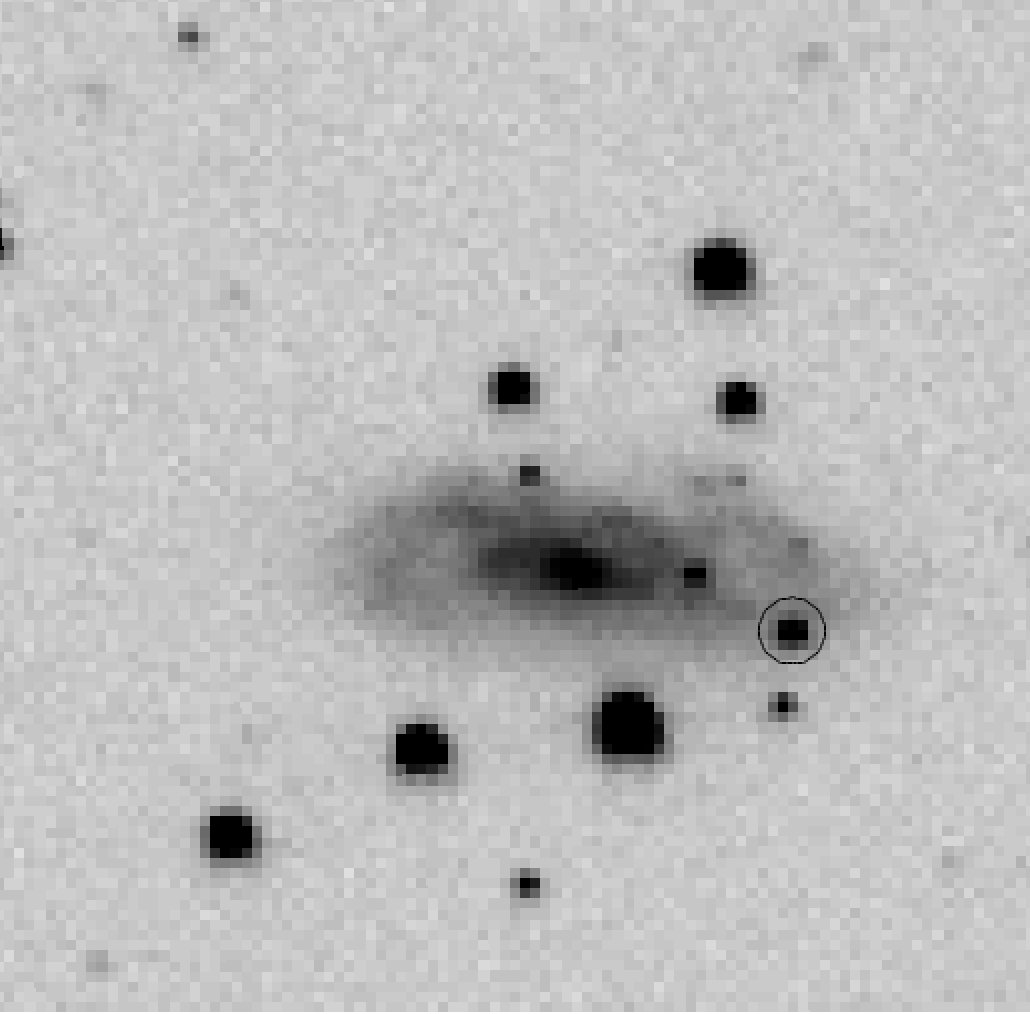}

\caption{Left panel: the PS1 $r$-band image\footnote{http://ps1images.stsci.edu} of UGC 85, the host galaxy of the supernova SN~2018\,fif. Right panel: the P48 $r$-band image of SN~2018\,fif on September 4 2018, at 9:26:50.00 UT. The circle is centered on the SN position.}
\label{fig:host}
\end{center}
\end{figure*}

SN~2018\,fif was first detected on 2018 August 21 at 8:46 UT by the ZTF wide-field camera mounted on the $1.2$\,m Samuel Oschin Telescope (P48) at Palomar Observatory. ZTF images were processed and calibrated by the ZTF pipeline \citep{Masci2019}. A duty astronomer reviewing the ZTF alert stream \citep{Patterson2019} via the ZTF GROWTH Marshal \citep{Kasliwal2019} issued an internal alert, triggering follow-up with multiple telescopes, using the methodology of \citep{Gal-Yam2011}.
This event was reported by \cite{Fremling2018} and designated SN\,2018fif by the IAU Transient Server (TNS\footnote{https://wis-tns.weizmann.ac.il/}). The SN is associated with the $B=14.5$ mag galaxy UGC 85 \citep{Falco1999}, shown in Figure~\ref{fig:host}. The coordinates of the object, measured in the ZTF images are $\alpha=00^h09^m26^s.55$, $\delta=+47^d21'14''.7$ (J2000.0). The redshift $z= 0.017189$ and the distance modulus $\mu=34.31\,\rm{mag}$ were obtained from the NASA/IPAC Extragalactic Database (NED) and the extinction was deduced from \cite{Schlafly2011} and using the extinction curves of \cite{Cardelli1989}. These parameters are summarized in Table~\ref{table:param}.

Previous ZTF observations were obtained in the months prior to the SN explosion and the most recent non-detection was on 2018 August 20 at 9:37:26.40\,UT, i.e. less than 24 hours before the first detection. We present a derivation of the explosion epoch in \S~\ref{sec:analysis_photo}. 

\begin{deluxetable}{lr}

\tablecolumns{2}
\tablecaption{Basic parameters of SN~2018\,fif}
\label{table:param}
\tablewidth{0pt}
\tablehead{\colhead{Parameter}&\colhead{Value}}
\startdata
Right ascension $\alpha$ ($J2000$)& $2.360644$\,deg \\
Declination $\delta$ ($J2000$)& $47.354093$\,deg\\ 
Redshift $z$&$z =0.017189$\\
Distance modulus $\mu$&$34.31$\,mag\\ 
Galactic extinction $E_{B-V}$&$0.10$\,mag  
\enddata
\end{deluxetable}
\subsection{Photometry}

\begin{figure*}
\begin{center}
\includegraphics[scale=.80]{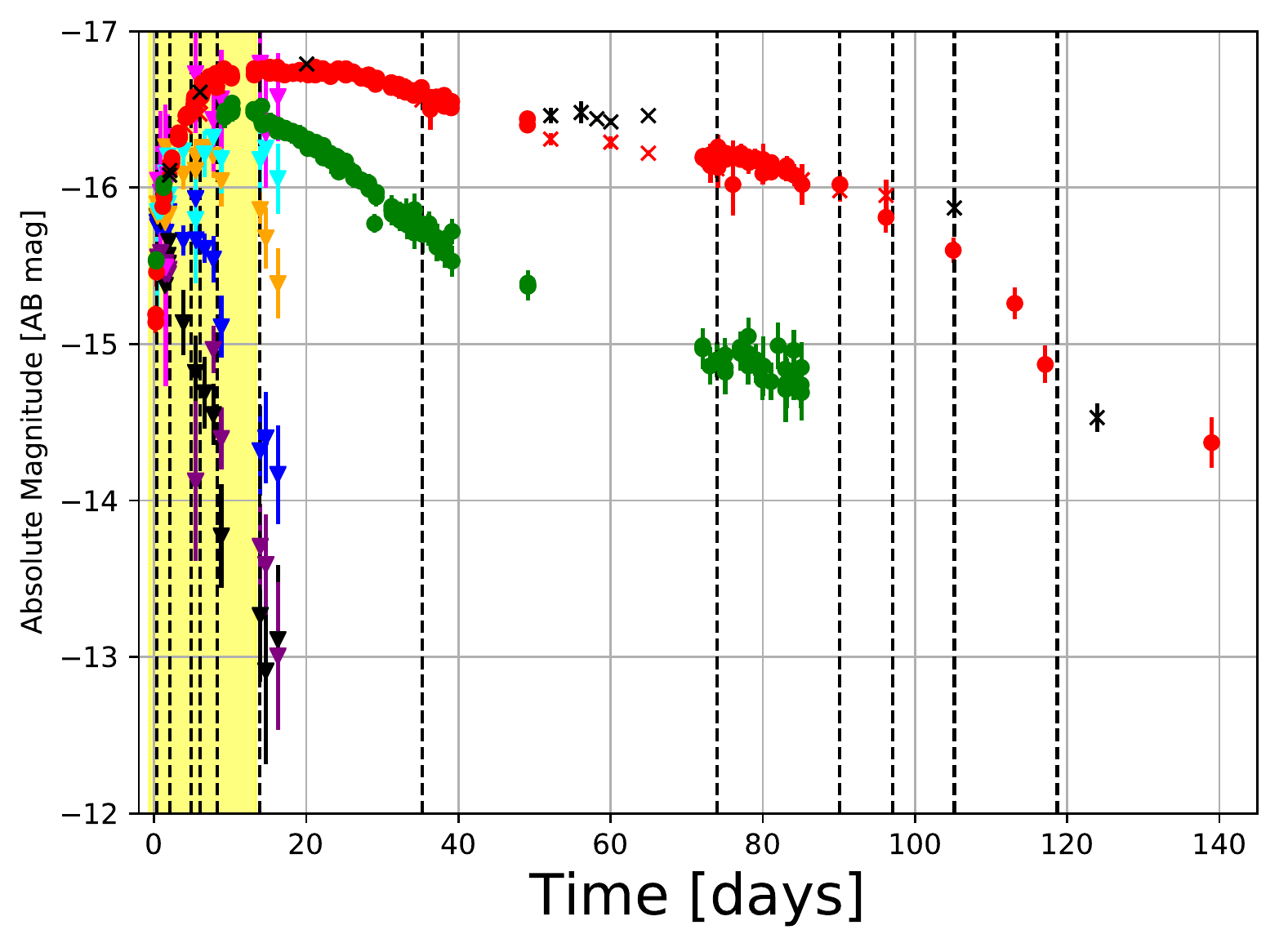}
\includegraphics[scale=.80]{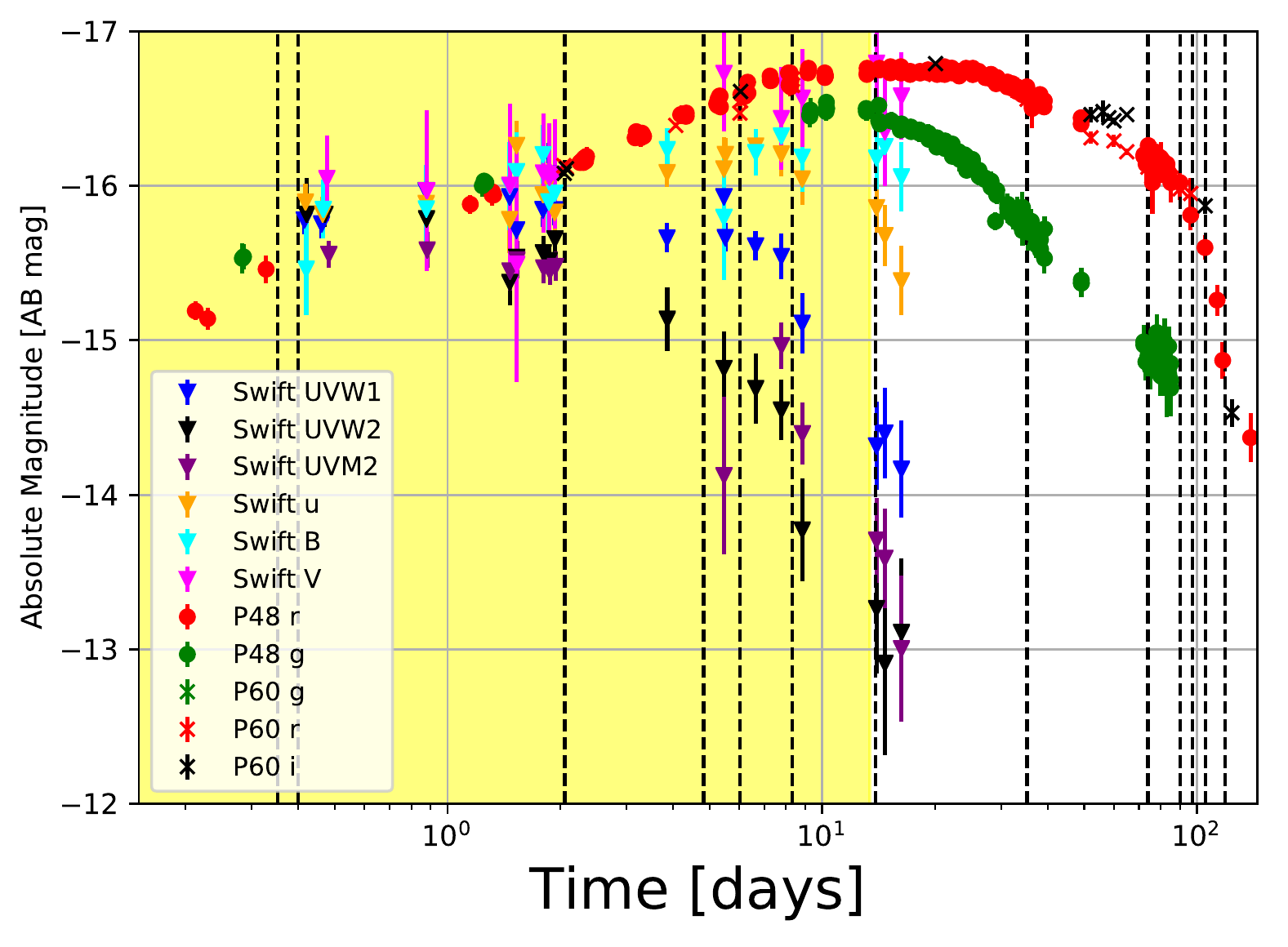}
\caption{The light curve of SN~2018\,fif in linear (top panel) and logarithmic space (lower panel). Time is shown relative to the estimated epoch at which the extrapolated light curve (Equation~\ref{eq:tref}) turns to zero: $t_{0}=2458351.6537$, as derived in \S~\ref{sec:analysis_photo}. Black dashed lines indicate dates at which spectroscopic data exist. The yellow background indicates the validity domain of the M20 best fit model:$[0.062,14.107]\,\rm days$ relative to the model explosion epoch $t_{\rm ref}$, i.e. $[-0.641,13.403]\,\rm days$ relative to $t_{0}$}.
\label{fig:lc}
\end{center}
\end{figure*}

SN~2018\,fif was photometrically followed in multiple bands for $\sim 5$ months. 
Light curves are shown in Figure~\ref{fig:lc}. The photometry is reported in Table~\ref{table:photo} and is electronically available from the Weizmann Interactive Supernova data REPository\footnote{https://wiserep.weizmann.ac.il} (WISeREP, \citealt{Yaron2012}).

\textit{Swift} \citep{Gehrels2004} observations of the SN~2018\,fif field started on 2018 August 21 and 11 observations were obtained with a cadence of $\sim1$ day. 

Observations from P48 were obtained using the ZTF mosaic camera composed of 16 $6$K$\times 6$K CCDs (e.g. \citealt{Bellm2015}) through SDSS $r$-band and $g$-band filters. Data were obtained with a cadence of $3$ to $6$ observations per day, to a limiting magnitude of $\rm{R}\approx20.5\,\rm mag [AB]$. ZTF data were reduced using the ZTF photometric pipeline \citep{Masci2019} employing the optimal image subtraction algorithm of \cite{Zackay2016}. 

Observations from the robotic $1.52$\,m telescope at Palomar (P60; \citealt{Cenko2016}) were obtained using the rainbow camera arm of the SED Machine spectrograph \citep{Blagorodnova2018}, equipped with a $2048\times2048$-pixel CCD camera and $g'$, $r'$, and $i'$ SDSS filters. P60 data were reduced using the FPipe pipeline \citep{Fremling2016}.

The UVOT data were retrieved from the NASA Swift Data Archive\footnote{\href{https://heasarc.gsfc.nasa.gov/cgi-bin/W3Browse/swift.pl}{ https://heasarc.gsfc.nasa.gov/cgi-bin/W3Browse/swift.pl }} and reduced using standard software distributed with HEAsoft version 6.26 \footnote{\href{https://heasarc.nasa.gov/lheasoft/}{ https://heasarc.nasa.gov/lheasoft/}}. Photometry was measured using the FTOOLSs uvotimsum and uvotsource with a 3" circular aperture. To remove the host contribution, we obtained and coadded two final epoch in all broad-band filters and built a host template using uvotimsum and uvotsource with the same aperture used for the transient.

\begin{deluxetable}{llll}
\tablecaption{Photometric observations of SN~2018\,fif}
\tablecolumns{4}
\tablewidth{0pt}
\tablehead{
\colhead{Epoch}&\colhead{Mag}&\colhead{Flux}&\colhead{Instrument}\\
\colhead{(jd)}&\colhead{(magAB)}&\colhead{($10^{-17}\rm erg/s/cm^2/\AA$)}&\colhead{}
}
\decimalcolnumbers
\startdata
$2458351.866$&$19.11\pm0.06$&$5.756\pm0.318$& P48/R\\
$2458351.937$&$18.78\pm0.10$&$15.10\pm1.391$ & P48/G\\
$2458353.697$&$18.18\pm0.02$&$15.263\pm0.281$ &  P60/r'\\
$2458353.699$&$18.17\pm0.03$&$26.563\pm0.734$ &  P60/g'\\
$2458353.7021$ &$18.23\pm0.02$&$9.907\pm0.183$ &P60/i'\\
$2458352.067$&$18.55\pm0.10$&$62.282\pm5.992$& \textit{Swift}/UVW1\\
$2458352.074$&$18.48\pm0.23$&$104.091\pm22.299$ &\textit{Swift}/UVW2\\
$2458352.132$&$18.71\pm0.09$&$70.281\pm6.024$ &\textit{Swift}/UVM2\\
$2458352.071$&$18.36\pm0.13$&$40.883\pm4.793$ &\textit{Swift}/u
\enddata
\tablecomments{This table is available in its entirety in machine-readable format in the online journal. A portion is shown here for guidance regarding its form and content.}
\label{table:photo}
\end{deluxetable}

\subsection{Spectroscopy}\label{sec:obs-spectroscopy}

Fifteen optical spectra of SN~2018\,fif were obtained using the telescopes and spectrographs listed in Table~\ref{table:obs}. All the observations were corrected for a galactic extinction of $E_{B-V}=0.10$\,mag, deduced from \cite{Schlafly2011} and using \cite{Cardelli1989} extinction curves. 

\begin{figure}
\begin{center}
\includegraphics[scale=.55]{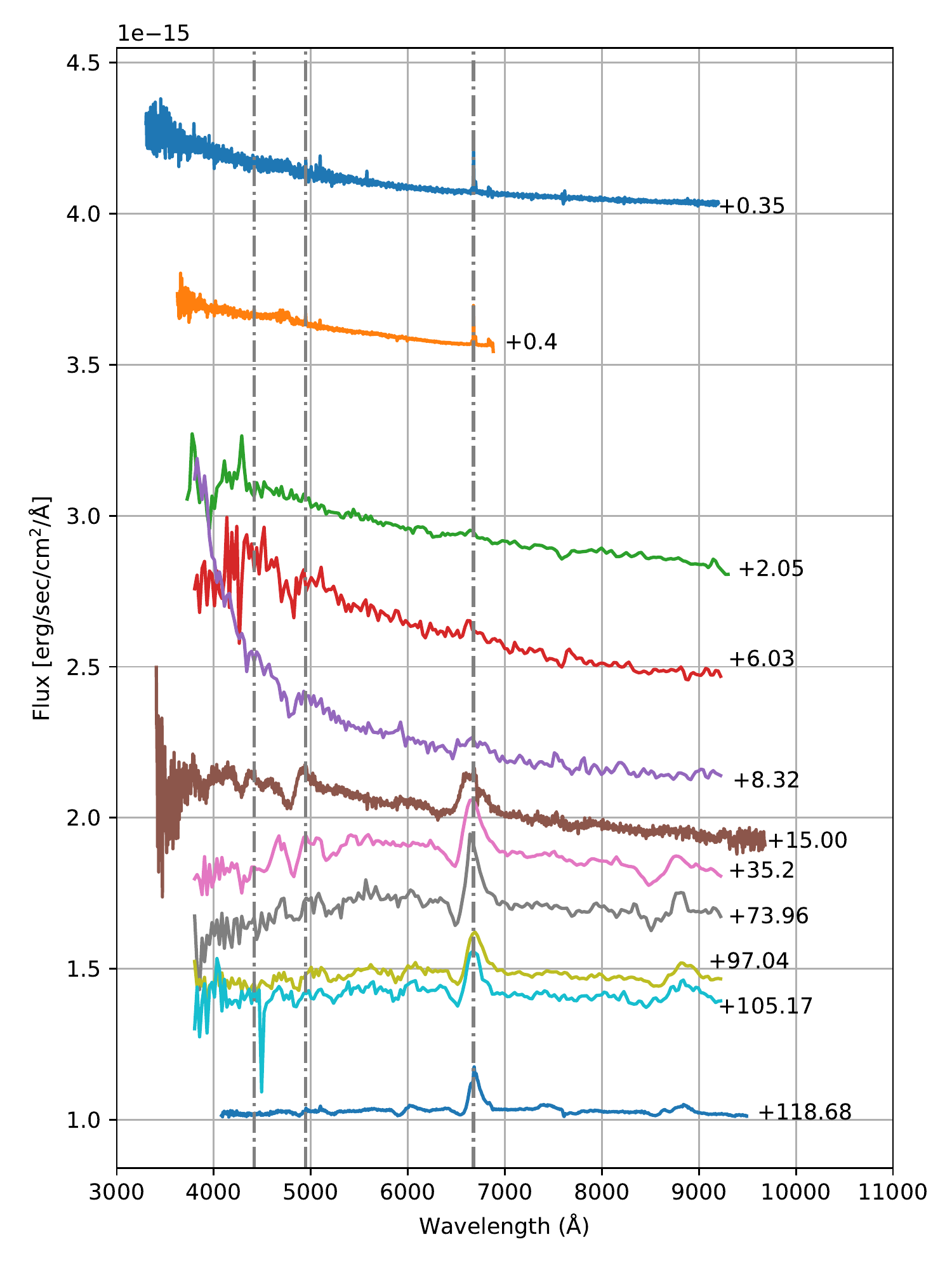}
\caption{The observed spectra of SN~2018\,fif. An offset was applied for easier visualization. Dashed lines indicate the redshifted emission lines for the Balmer series up to H$\gamma$. The phase is shown relative to the estimated epoch at which the extrapolated r-band light curve (based on Equation~\ref{eq:tref}) turns to zero: $t_{0}=2458351.6537$ (2018 August 21), as derived in \S~\ref{sec:analysis_photo}.}
\label{fig:spectra}
\end{center}
\end{figure}

 Following standard spectroscopic reduction, all spectra were scaled so that their synthetic photometry matches contemporaneous P48 $r$-band value. 
All spectra are shown in Figure~\ref{fig:spectra} and are available via WISeREP.

\begin{deluxetable*}{llccccrl}
\tablecaption{Spectroscopic observations of SN~2018\,fif}
\tablecolumns{7}
\tablewidth{0pt}
\tablehead{\colhead{Date}  &  \colhead{Phase}  &  \colhead{Facility [Ref]}  &  \colhead{Exp. T}  &  \colhead{Grism/Grating} & \colhead{Slit} &  \colhead{R}  &  \colhead{Range} \\
\colhead{(2018UT)}  &  \colhead{(days)}  &  \colhead{}  &  \colhead{(s)}  &  \colhead{}  &  \colhead{$('')$}  &  \colhead{}  &  \colhead{(\AA)}}
\startdata
08-21 12:08:32  &  $+$0.35   &  P200/DBSP [1]  & 900  & 600/4000$+$316/7500  &  1.5 & -  & 3310$-$9190 \\
08-21 12:08:01  &  $+$0.35   &  P60/SEDM [3],[4]   & 2430  & IFU  &  & $\sim$100  & 3700$-$9300  \\
08-21 12:25:04  &  $+$0.40   &  Gemini N/GMOS [2]  &   900$\times$4  & B600  &  1.0  & 1688 & 3630$-$6870 \\
08-23 04:59:25  &  $+$2.05   &  P60/SEDM [3],[4]   &  1440  & IFU &  & $\sim$100  & 3700$-$9300  \\
08-25 23:25:40  &  $+$4.82   &  LT/SPRAT [5]   &  300   &   &  1.8  & 350        & 4020$-$7960  \\
08-27 04:22:22  &  $+$6.03   &  P60/SEDM [3],[4]   &  1440  & IFU &  & $\sim$100  & 3780$-$9220  \\
08-29 11:22:34  &  $+$8.32   &  P60/SEDM [3],[4]   &  1440  & IFU &  & $\sim$100  & 3780$-$9200  \\
09-05 03:46:42  &  $+$15.00  &  NOT/ALFOSC     &   1800  & Grism 4 &  1.0  &  360  & 3410$-$9670  \\
09-25 08:33:17  &  $+$35.20  &  P60/SEDM [3],[4]   &  1440  & IFU &  &  $\sim$100  & 3780$-$9220  \\
11-03 02:50:19  &  $+$73.96  &  P60/SEDM [3],[4]   &  1600  & IFU &  &  $\sim$100  & 3780$-$9220  \\
11-14 07:53:52  &  $+$85.17  &  P60/SEDM [3],[4]   &  1200  & IFU &  &  $\sim$100  & 3780$-$9220  \\
11-19 06:25:58  &  $+$90.11  &  P60/SEDM [3],[4]   &  1200  & IFU &  &  $\sim$100  & 3780$-$9220  \\
11-26 04:39:18  &  $+$97.04  &  P60/SEDM [3],[4]   &  1200  & IFU &  &  $\sim$100  & 3780$-$9220  \\
12-04 07:48:03  &  $+$105.17 &  P60/SEDM [3],[4]   &  1200  & IFU &  & $\sim$100  & 3780$-$9220  \\
12-17 20:01:45  &  $+$118.68 &  WHT/ACAM  [6]   &  1500$\times$2  & V400  & 1.0 & 450  & 4080$-$9480  \\
\enddata
\tablecomments{
[1]:\cite{Oke1982}; [2]:\cite{Oke1994}; [3]:\cite{Blagorodnova2018}; {\color{blue}[4]:\cite{Rigault2019}}; [5]:\cite{Steele2004}; [6]:\cite{Benn2008}}%
\label{table:obs}
\end{deluxetable*}


\section{Analysis}
\subsection{Epoch of first light}\label{sec:analysis_photo}

We fitted the P48 r-band rising flux during the first week with a function of the form
\begin{equation}\label{eq:tref}
f = a(t-t_0)^n\,,
\end{equation}
where $t_0$ is the time of zero flux. This allowed us to estimate the epoch at which the extrapolated r-band light curve turns to zero, which is used throughout this paper as the reference time $t_0 (MJD) = 58351.1537_{-0.0903}^{+0.0356}$ (2018 Aug 21 at 03:41:19.680 UTC, $0.2$ days before the first r-band detection). 

\subsection{Black body temperature and radius}\label{sec:peculiar_r}

\begin{figure*}\label{fig:evo}
\begin{center}
\includegraphics[scale=.50]{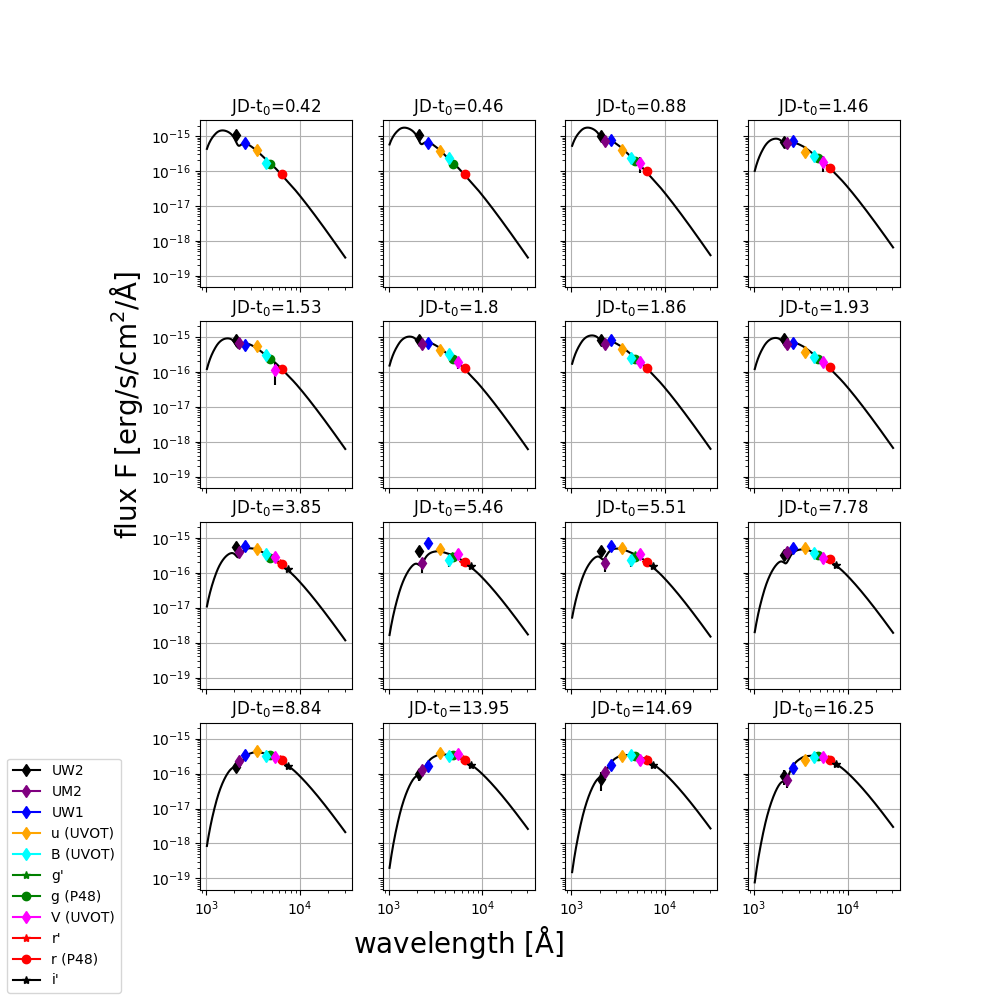}
\caption{Black body fits to Swift/UVOT and optical photometry for SN\,2018\,fif.  Using the {\tt PhotoFit} tool (released in the appendix), photometric points were interpolated to a common epoch (UVOT epochs), and the errors at the interpolated points were computed with Monte Carlo Markov-chain simulations.}
\label{fig:seds}
\end{center}
\end{figure*}

\begin{figure}\label{fig:evo}
\begin{center}
\includegraphics[scale=.43]{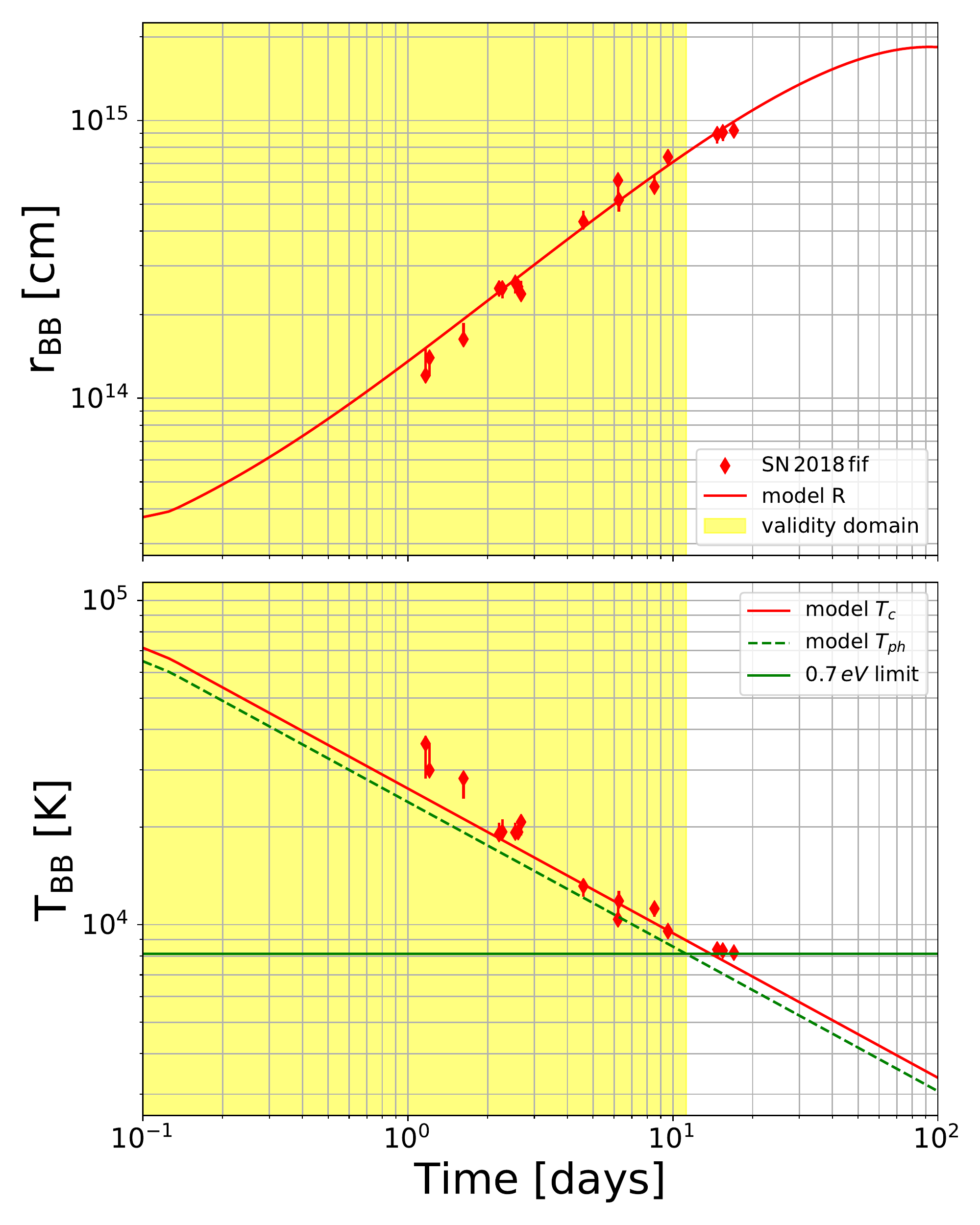}
\caption{The evolution in time of: (1) the radius (top panel) and (2) the temperature (lower panel) of a blackbody with the same radiation as SN\,2018\,fif (red). The points were obtained using the {\tt PhotoFit} code (released in the appendix).The reference time is the best-fit $t_{\rm exp}$. The yellow background indicates the temporal validity domain of the M20 best fit model: $[0.062,14.107]\,\rm days$  relative to $t_{\rm exp}$. The red continuous line indicates the radius $R$ and color temperature $T_{\rm col}$ predicted by M20 for the best fit model. The green dashed line indicates $T_{\rm ph,RW}$ (linked to $T_{\rm col}$ through $T_{\rm col}/T_{\rm ph,RW}=1.1[1.0]\pm 0.05$, see section~\ref{sec:SW2017}) and the continuous green line shows the $0.7\,eV$ temperature. The time at which $T_{\rm ph,RW}$ drops below $0.7\,eV$ defines the upper limit of the temporal validity window.}
\label{fig:evo_param_TR}
\end{center}
\end{figure}

Taking advantage of the multiple-band photometric coverage, we derived the temperature and radius of the black body that best fits the photometric data at each epoch after interpolating the various data sets to obtain data coverage at coinciding epochs, and deriving the errors at the interpolated points with Monte Carlo Markov-chain simulations. This was performed using the {\tt PhotoFit}\footnote{https://github.com/maayane/PhotoFit} tool, which is released in the appendix. The extinction $E_{\rm B-V}$ was implemented using the extinction curves by \cite{Cardelli1989} with $R_V=3.08$. The interpolated SEDs are shown in Figure~\ref{fig:seds}. The derived best-fit temperatures $T_{BB}$ and radii $r_{BB}$ are shown and compared to those derived for SN~2013\,fs in Figure ~\ref{fig:evo_param_TR}. 

\subsection{Bolometric light curve}\label{sec:bolo}
\begin{figure}
\begin{center}
\includegraphics[scale=.55]{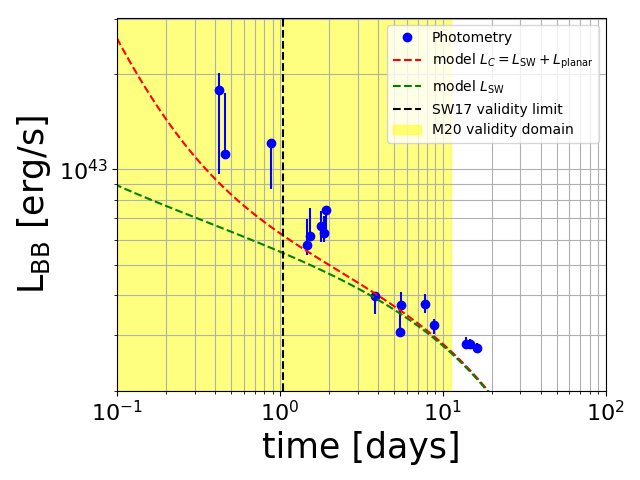}
\caption{The evolution in time of the bolometric luminosity of a blackbody with the same radiation as SN\,2018\,fif. The green and red dashed lines show the SW17 and M20 predictions, respectively. The yellow background indicates the validity domain of the M20 best fit model: $[0.062,14.107]\,\rm days$  relative to $t_{\rm exp}$ while the black dashed line shows the lower limit of the SW17 model temporal validity window for the same set of progenitor's parameters.}
\label{fig:evo_param_L}
\end{center}
\end{figure}

Based on the measurement of $r_{BB}$ and $T_{BB}$, we were able to derive the luminosity $L_{BB}=4\pi r_{BB}^2\sigma T_{BB}^4 $ of the blackbody fits, shown in Figure~\ref{fig:evo_param_L}. It is interesting to note that the bolometric peak occurs early on during the UV-dominated hot shock-cooling phase, well before the apparent peak at visible light.

\subsection{Spectroscopy}\label{sec:analysis_spectroscopy}

Figure~\ref{fig:spectra} shows the spectroscopic evolution of SN\,2018fif over 119\,d from its estimated explosion time. The sequence is quite typical for Type II SNe \citep{Gal-Yam2017}, initially showing blue, almost featureless spectra, with low-contrast Balmer lines emerging and becoming pronounced after about a week. The spectrum at phase 15.00\,d is typical for the early photospheric phase, with a relatively blue continuum and strong Balmer lines, with H$\alpha$ showing a strong emission component, H$\beta$ having a symmetric P-Cygni profile, and H$\gamma$ appearing only in absorption. The spectra continue to develop during the slowly declining light curve phase over several months, with the continuum emission growing redder and lines becoming stronger. The latest spectra approach the nebular phase and are dominated by a strong emission component of the H$\alpha$ line, emerging emission lines of Ca II (at $\lambda$\,7300\,\AA\ as well as the NIR triplet), weaker OI ($\lambda$\,7774\,\AA\ and a hint of $\lambda$\,6300\,\AA) and Na D. 

\begin{figure}
\includegraphics[scale=.25]{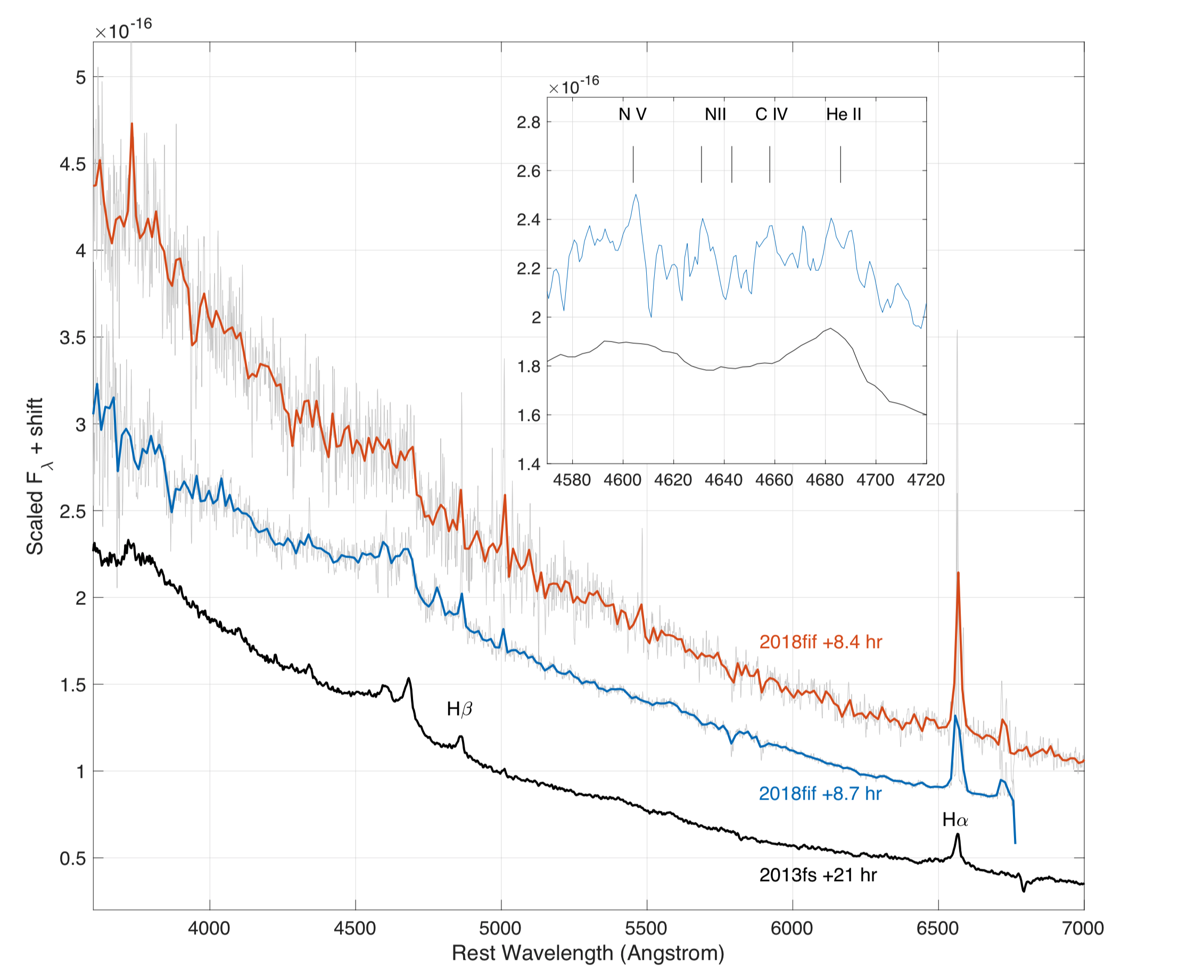}
\caption{Comparison of early spectra of SN\,2018\,fif (at 8.4 and 8.7\,hr) and SN\,2013\,fs (at 21\,hr; from \citealt{Yaron2017}). SN\,2018\,fif shows sharp, narrow Balmer lines lacking a broad electron-scattering base. A broad ledge around $4600\,\rm \AA$ indicates a likely blend of weak high-ionization lines, suggesting some CSM emission does exist in this event, though less than in SN\,2013\,fs, see text. }
\label{fig:compare_spec}
\end{figure}

Focusing on the earliest phase, 
in Figure~\ref{fig:compare_spec}, we show a comparison of the early spectra of SN\,2018\,fif (P200/DBSP and Gemini-N/GMOS at +8.4 and +8.7 hrs from the estimated explosion time, respectively) with the +21 hr NOT/ALFOSC spectrum of SN~2013\,fs \citep{Yaron2017}, which is most similar to our data. We note that earlier spectra of SN~2013\,fs at similar phase to those of SN\,2018fif ($6-10$\,h after explosion) are dominated by very strong emission lines of OIV and HeII that are not seen in this case. 

In the spectrum of SN~2013\,fs, the hydrogen Balmer lines show a broadened base and characteristic electron-scattering wings that are a measure of the electron density in the CSM. The spectra of SN~2018\,fif do not show such electron-scattering signatures, even at a much earlier time, and the narrow emission lines seem to arise only from host galaxy emission, with similar profiles to other host lines (such as NII and SII, evident right next to the H$\alpha$ line).
A signature of some CSM interaction may appear in the blue part of the spectrum, in a ledge-shaped emission bump near $\lambda$\,4600\,\AA. This shape is similar to that seen in the SN 2013fs spectrum, though the sharp emission spikes (in particular of HeII $\lambda$\,4686\,\AA) are less well defined. The inset in Figure~\ref{fig:compare_spec} shows a zoom-in of the elevated region around the HeII $\lambda$\,4686\,\AA, emission line for both the SN~2018\,fif +8.7\,hr and the SN~2013\,fs +21\,hr spectra. Possible emission lines that may contribute to this elevated emission region include NV $\lambda$\,4604, NII $\lambda$\, 4631,$\lambda$\,4643 and CIV $\lambda$\,4658\,\AA. Although these identifications are not certain (since they are based on single lines that are only marginally above the noise level), it appears likely that a  blend of high-ionization lines is responsible for the elevated emission above the blue continuum. 

The difference between the spectra of SN~2013\,fs and SN~2018\,fif at $\sim$ 8 hrs, and in particular the fact that SN~2013\,fs showed much stronger lines of higher ionization species at similar epochs, suggests that the progenitor of SN~2018\,fif was surrounded by less nearby CSM than the progenitor of SN~2013\,fs. The lack of strong high-ionization lines in the spectra of SN\,2018fif, as well as the sharp profiles of the Balmer lines that show no evidence of electron-scattering wings, suggest that the CSM that did surround the progenitor of SN~2018\,fif was likely less dense than in the case of SN~2013\,fs.

\section{Shock cooling and progenitor model}\label{sec:modeling}

\subsection{The model}\label{sec:modeling-SW}

In order to model the multiple-bands emission from SN~2018\,fif, we used the model by \cite{SW2017}, an extension of the model derived in \cite{Rabinak2011}. In the following, the abbreviations "SW17" and "RW11" are used to refer to the models. We summarize below the main conclusions of these two models. Both hold for temperatures $>0.7\,\rm{eV}$, the limit above which Hydrogen is fully ionized, where recombination effects can be neglected and the approximation of constant opacity holds. We emphasize that the results presented here depend on the assumptions adopted by the SW17 analytical model we use, and that other approaches - in particular using hydrodynamical codes - exist and could be used for modeling our observations.

\subsubsection{The \cite{Rabinak2011} model}

\cite{Rabinak2011} explored the domain of times when the emission originates from a thin shell of mass i.e. the radius of the photosphere is close to the radius of the stellar surface. The post-breakout time-evolution of the photospheric temperature and bolometric luminosity, are given below (see also Equation (4) of \citealt{SW2017}), where the prefactors correspond to different power law indexes for the density profiles close to the stellar surface (i.e. at radii $r$ such as  $\delta\equiv (R-r)/R <<1$) $\rho\propto\delta^n$ with $n=3/2[3]$ for convective[radiative] polytropic envelopes (see Equation (1) in \citealt{SW2017}):

\begin{equation}\label{eq:T_ph_RW}
T_{\rm ph,RW}=1.61[1.69]\left( \frac{v^2_{\rm s*,8.5}t_{\rm d}^2}{f_\rho M_0\kappa_{0.34}}\right)^{\epsilon_1}\frac{R_{13}^{1/4}}{\kappa_{0.34}^{1/4}}t_{\rm d}^{-1/2}\, \rm{eV}\,,
\end{equation}

\begin{equation}
L_{\rm RW}=2.0[2.1]\times 10^{42}\left( \frac{v_{\rm s*,8.5}t_{\rm d}^2}{f_\rho M_0\kappa_{0.34}}\right)^{-\epsilon_2}\frac{v^2_{\rm s*,8.5}R_{13}}{\kappa_{0.34}}\, \rm{erg/s}\,,
\end{equation}

\noindent where $\kappa=0.34\kappa_{0.34}\, \rm cm^2\,g^{-1}$, $v_{\rm s*}=10^{8.5}\, v_{\rm s*,8.5}$, $M=M_0\,M_\odot$, $R=10^{13}\,R_{13}\rm cm$, $\epsilon_1=0.027[0.0.016]$ and $\epsilon_2=0.086[0.175]$. $M$ is the mass of the ejecta, $f_\rho$ is a numerical factor of order unity describing the inner structure of the envelope,  $t_{\rm d}$ is the time from explosion in days, and $v_{\rm s*}$ is a measure of the shock velocity $v_{\rm sh}$: in regions close to the stellar surface, $v_{\rm sh}$ is linked to $v_{\rm s*}$ through \citep{Gandelman1956,Sakurai1960}
\begin{equation}\label{eq:sw2}
v_{\rm sh}=v_{\rm s*}\delta^{-\beta n}\,,
\end{equation}
where $\beta=0.191[0.186]$, $n=3/2[3]$ and where $v_{\rm s*}$ only depends on E, M (the ejecta energy and mass) and $f_{\rho}$ \citep{Matzner1999}:
\begin{equation}\label{eq:sw3}
v_{\rm s*}\approx1.05f_{\rho}^{-\beta}\sqrt{E/M}\,,
\end{equation}

The RW11 model holds during a limited temporal range. The upper limit on this range,

\begin{equation}\label{eq:thin}
t<3f_{\rho}^{-0.1}\frac{\sqrt{\kappa_{0.34}M_0}}{v_{\rm s*,8.5}}\,\rm{days}
\end{equation}
follows from the requirement that the emitting shell carry a small fraction of the ejecta mass. The lower limit

\begin{equation}\label{eq:SW_early_validity}
t>0.2\frac{R_{13}}{v_{\rm s* 8.5}}max\left[0.5,\frac{R_{13}^{0.4}}{(f_{\rho}\kappa_{0.34}M_0)^{0.2}v_{\rm s* 8.5}^{0.7}}\right]
\end{equation}
comes from two different requirements: (1) The photosphere must have penetrated beyond the thickness at which the initial breakout happens (see equation (16) of RW11) and (2) Expansion must be significant enough so that the ejecta are no longer planar and have become spherical \citep{WaxmanKatz2016}; this last requirement was added to the model of \cite{SW2017}.

\subsubsection{The \cite{SW2017} model}\label{sec:SW2017}

\cite{SW2017} extended the RW11 description to later times, when the photosphere has penetrated more deeply into the envelope, but is still close enough to the surface so that the emission is still weakly dependent on the inner structure of the envelope. As radiation originates from inner regions, the self-similar description of the shock-wave \citep{Gandelman1956,Sakurai1960}, one of the key ingredients of the RW11 model, does not hold anymore. This results in a suppression of the bolometric luminosity that can be approximated by (equation (14) of \citealt{SW2017}):
\begin{equation}\label{eq:L_bolo}
L/L_{\rm RW}=A\exp\left[-\left(\frac{at}{t_{\rm tr}}\right)^\alpha\right]\,,
\end{equation}

where $A=0.94[0.79]$, $a=1.67[4.57]$ and $\alpha=0.8[0.73]$ for convective[radiative] envelopes. The thin shell requirement (Equation~\ref{eq:thin}) is relaxed, and the new upper limit of the valid time range is dictated by the requirement of constant opacity:

\begin{equation}
t<min(t_{\rm tr}/a,t_{\rm{T<0.7}})\,,
\end{equation}
where $t_{\rm tr}$ is the time beyond which the envelope becomes transparent, and $t_{\rm{T<0.7}}$ is the time when T drops below $0.7\,\rm eV$ and recombination leads to a decrease of the opacity.

The observed flux, for a SN at luminosity distance $D$ and redshift $z$  is given by

\begin{equation}\label{eq:fluxes}
f_{\lambda}(\lambda,t)=\frac{L(t)}{4\pi D^2\sigma T_{col,z}^4} B_\lambda(\lambda,T_{\rm col,z})
\end{equation}


where $T_{\rm col,z}=T_{\rm col}/(z+1)$ is the temperature of a blackbody with intrinsic temperature $T_{\rm col}$, observed at redshift $z$, $T_{\rm col}/T_{\rm ph,RW}=1.1[1.0]\pm 0.05$ for convective[radiative] envelopes, $L$ is the bolometric luminosity given in equation~\ref{eq:L_bolo} and $B_\lambda$ is the Planck function 
\begin{equation}
    B_\lambda=\frac{2\pi hc^2}{\lambda^5}\frac{1}{e^{\frac{hc}{\lambda k_BT}}-1}
\end{equation}

\subsubsection{The Morag, Sapir \& Waxman extension to early times}\label{sec:MSW}
Morag et. al. (in preparation) further extend the \citealt{SW2017} prescription to account for the transition from a planar shock breakout to a spherical self-similar motion of the ejecta. The new  prescription is still described by eq. \ref{eq:fluxes}, but with a modified luminosity and temperature.
The new composite luminosity $L_{C}$ is given by
\begin{equation}
    L_{C} = L_{planar} + L_{SW}
\end{equation}
where $L_{SW}$ is given in eq. \ref{eq:L_bolo}, and 
\begin{equation}
L_{planar}=2.974\times10^{42}\frac{R_{13}^{0.462}v_{\star,85}^{0.602}}{\left(f_{\rho}M_{0}\kappa_{34}\right)^{0.0643}}\,\frac{R_{13}^{2}}{\kappa_{34}}t_{h}^{-4/3}\:\text{erg}\,\text{s}^{-1}.
\end{equation}

$L_{planar}$ is the planar stage post-breakout luminosity as given by eq. 23 in \cite{sapir_katz_waxman_I_2011}, cast in terms of SW17 variables. Note that $t_{h}$ in hours has replaced $t_{d}$ in days.
Likewise, the color temperature is given by

\begin{equation}
T_{C}=f_{T}\min\left[T_{planar},T_{\rm ph,RW}\right]
\end{equation}

where $T_{\rm ph,RW}$ is the photosphere temperature given in eq. \ref{eq:T_ph_RW}, not including the 1.1 SW17 color factor. Meanwhile,

\begin{equation}
T_{planar}=6.937\,\frac{R_{13}^{0.1155}v_{\star,85}^{0.1506}}{f_{\rho}^{0.01609}M_{0}^{0.01609}\kappa_{34}^{0.2661}}t_{h}^{-1/3}\,\text{eV}
\end{equation}

is the post-breakout temperature during the planar stage, as given by \cite{sapir_katz_waxman_III_2013} (sec. 3.2). A color factor of $f_{T}=1.1$ is still an appropriate choice, as calibrated against a set of grey diffusion simulations for a wide range of progenitors and explosion energies.

The new emission accounts for light travel-time effects (\cite{katz_sapir_waxman_II_2012}) and has the important advantage that it agrees with grey diffusion simulations as early as $t=3R_{\star}/c=17R_{13}\,{\rm min} $ after breakout. Thus, we are able to ignore the early SW17 validity times (eq. \ref{eq:SW_early_validity}) and include all the early data points immediately following breakout in our fit.

\subsection{The {\tt SOPRANOS} algorithm}\label{sec:modeling-code}

The main difficulty in implementing the SW17 model is that the temporal validity domain of the model depends on the parameters of the model themselves. In other words, different combinations of the model's parameters correspond to different data to fit \citep{Rubin2017}. One way to cope with this difficulty is to fit the data for a chosen range of times, and to retrospectively assess whether the solution is valid in this temporal window. This approach, which was taken e.g. by \cite{Valenti2014,Bose2015,Rubin2016} and \cite{Hosseinzadeh2019}, is not fully satisfactory for several reasons: (1) it may limit the explored area in the parameter space, since this area is pre-defined by the choice of the data temporal window and (2) it makes it impossible to make a fair comparison between models, as the goodness of a model should be judged on nothing more or less than its specific validity range: a good model fits the data on its {\it entire} validity range and {\it only} on its validity range.
It is clear that the best-fit model (and hence deduced progenitor parameters) may depend on the arbitrary choice of pre-defined data modeled, which is not a good result.


Here, we adopt a self-consistent approach and build an algorithm to find models that fit well the data included in their entire range of validity. In this sense, our approach is similar to the one adopted by \cite{Rubin2017}. The {\tt SOPRANOS} algorithm (ShOck cooling modeling with saPiR \& wAxman model by gANOt \& SOumagnac, Ganot et al. in preparation) is available in two versions: {\tt SOPRANOS-grid}, written in {\tt matlab} and {\tt SOPRANOS-nested}, written in {\tt python} (Ganot et al., in preparation). The steps of {\tt SOPRANOS-grid} are as follows:
\begin{itemize}
    \item we build a 6-dimensional grid of parameters $\rm \{R,v_{\rm s*,8.5 },t_{\rm ref},M,f_{\rho},E_{\rm B-V}\}$: a given point in the grid (indexed e.g. $j$, for clarity) corresponds to a model $\mathcal{M}_j$;
    \item we calculate, for each point in the grid, the time-validity domain, and deduce from it the set of $N_j$ data points $\{x_i,y_i\}_{i\in [1,N_j]}$ (with uncertainties $\sigma_{y_i}$ on the $y_i$ values) to be taken into account in the fit of model $\mathcal{M}_j$ to the data;
    \item we calculate a probability for each point in the grid, using
    \begin{equation}\label{eq:pdf}
        P_j=PDF(\chi_j^2,\nu_j)\,,
    \end{equation}
    where $\nu_j$ is the number of degrees of freedom (this number varies between models, as the validity domain - and hence the number of points included in the data - varies), $\chi_j^2$ is the chi-square statistic of the fit, for the model $\mathcal{M}_j$ 
    \begin{equation}
        \chi_j^2=\sum_{i=1}^{N_j} \frac{(y_i-\mathcal{M}_j(x_i))^2}{\sigma_{y_i}^2}\,
    \end{equation}
    
    and $PDF$ is the chi-squared probability distribution function.
\end{itemize}

The output of this procedure is a grid of probabilities, which we can compare to each other to find the most probable model. 

The fluxes $\mathcal{M}_j(x_i))$ are calculated based on equation~\ref{eq:fluxes}. The extinction $E_{\rm B-V}$, a free parameter of the model, is applied to the full spectrum using the extinction curves by \cite{Cardelli1989} with $R_V=3.08$. Synthetic photometry is then computed using the {\tt pyphot} algorithm\footnote{http://mfouesneau.github.io/docs/pyphot/} (Fouesneau, in preparation), to convert the monochromatic fluxes $f_{\lambda}$ into band fluxes.

The second version of the {\tt SOPRANOS} algorithm, {\tt SOPRANOS-nested}, uses the model probability defined in equation~\ref{eq:pdf} as the input of the nested sampling algorithm {\tt dynesty} \citep{Skilling2004,Skilling2006,Higson2019,Speagle2020}. 

\begin{figure}
\includegraphics[scale=.50]{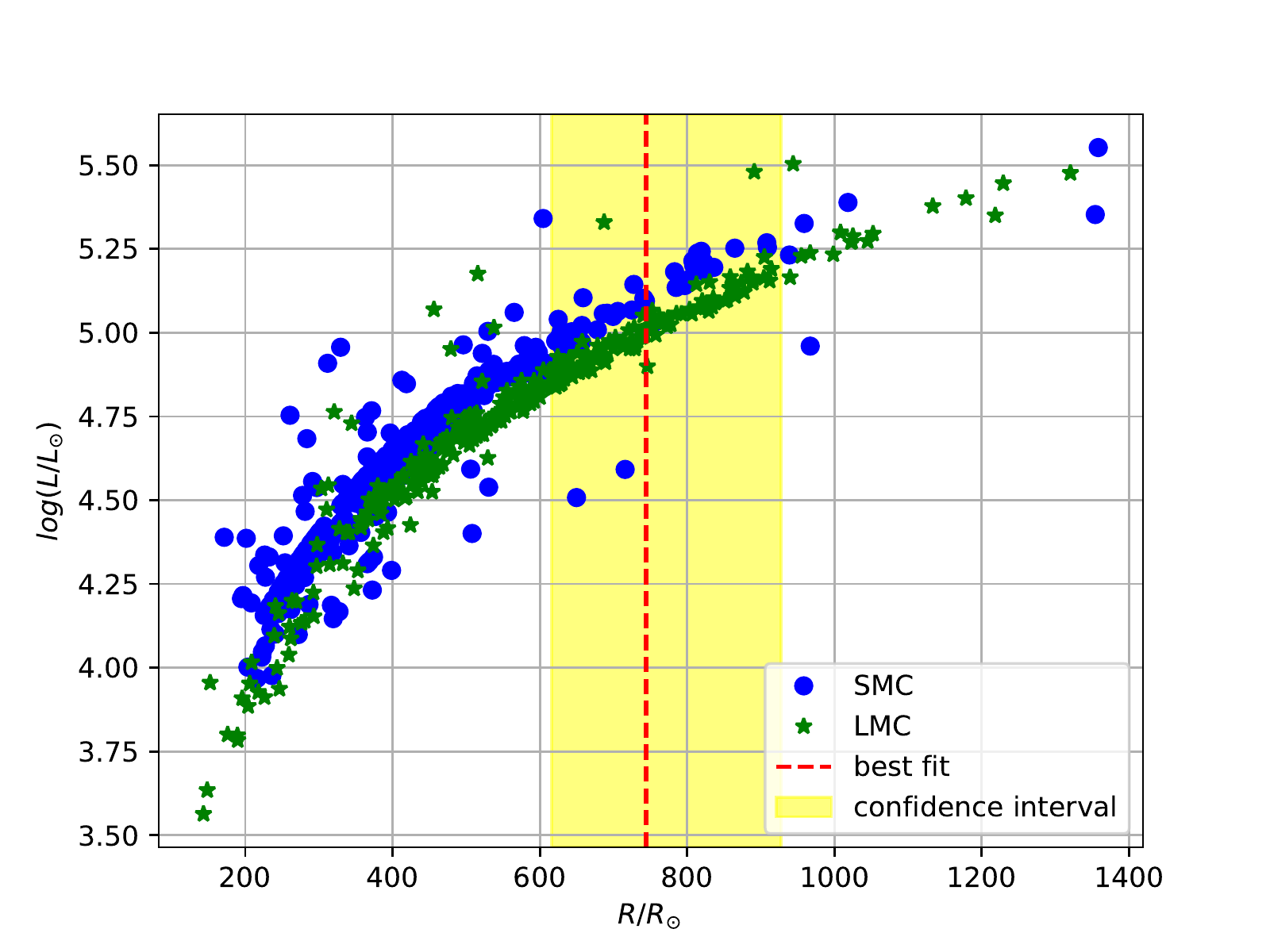}
\caption{Radii and luminosities of the stars in the small and large Magellanic Clouds, derived from the effective temperatures and luminosities published by \cite{Davies2018}. The dashed line shows the best-fit solution for SN\,2018\,fif and the yellow background shows the confidence interval.}
\label{fig:L_R_Davies}
\end{figure}

In both cases, we apply the following flat priors for the six parameters of our model: $R\in [200,1500]$, $v_{\rm s*,8.5}\in [0.3,1.5]$, $M\in [2,25]$, $f_{\rm \rho}\in [\sqrt{1/3},\sqrt{10}]$ \citep{SW2017}, $t_{\rm ref}\in [2458347.5,t_0]$, $E_{\rm B-V}\in [0.1,0.35]$. The prior on the radius $R$ was chosen to reflect the bulk of current measurements (\citealt{Davies2018}; see Figure~\ref{fig:L_R_Davies} and section \ref{sec:discussion} for a discussion on higher radii). The prior on $f_{\rm \rho}\in [\sqrt{1/3},\sqrt{10}]$ corresponds to the range used in the model by \cite{SW2017}. The choice of priors for $t_{\rm ref}$, $v_{\rm s*,8.5}$ and $E_{\rm B-V}\in [0.1,0.35]$ is the result of an iterative process (coarse to fine grid) aiming at finding the relevant location in the parameter space while limiting the memory use and running time. In all our analysis, we use $\kappa=0.34\, \rm cm^2\,g^{-1}$ and assume a convective envelope for the progenitor. The NOT spectrum taken on September 5, 2018 ($JD=246366.5$, phase +15.00, see Figure~\ref{fig:spectra}), showing strong hydrogen lines with  p-cygni profiles, gives an upper limit on the time of recombination, beyond which the SW17 and M20 models are not valid. In practice, recombination is likely to have occurred several days before the emergence of such strong p-cygni profiles, but in the absence of earlier spectra showing such patterns, this gives a conservative prior on the temporal validity of the models, which we also implement in our algorithm.

Note that our approach is similar to the one by \cite{Rubin2017}, in the sense that it is self-consistent and takes care of the time-validity issue. However, the strategy adopted to compare and discriminate between models (equation~\ref{eq:pdf}) is different. Another difference is the use of the M20 model, which allows us to ignore the SW17 lower limit of the time-validity window and use all the early data.


\subsection{Results}\label{sec:modeling-results}

\begin{figure}
\begin{center}
\includegraphics[scale=.55]{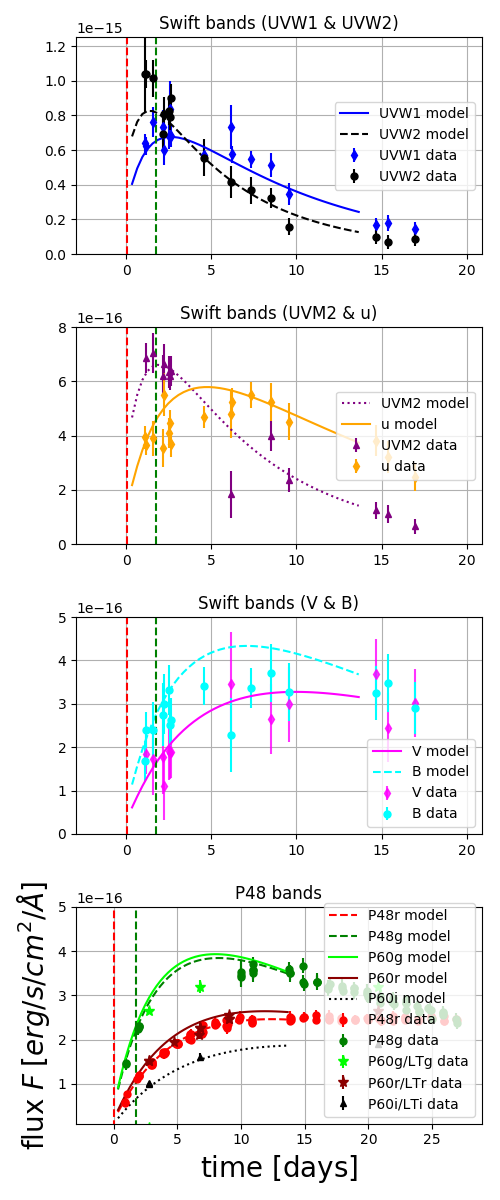}
\caption{Best fit Morag, Sapir \& Waxman model ($\chi^2/dof=1.69$) superimposed with the photometric data of SN\,2018\,fif. The dashed line indicates the lower limit of the temparal validity window for the SW17 (green) and M20 (red) models}. The reference date is $t_{\rm ref}$, the explosion epoch predicted by our model.
\label{fig:lc_versus_model}
\end{center}
\end{figure}

\begin{figure*}
\includegraphics[scale=.50]{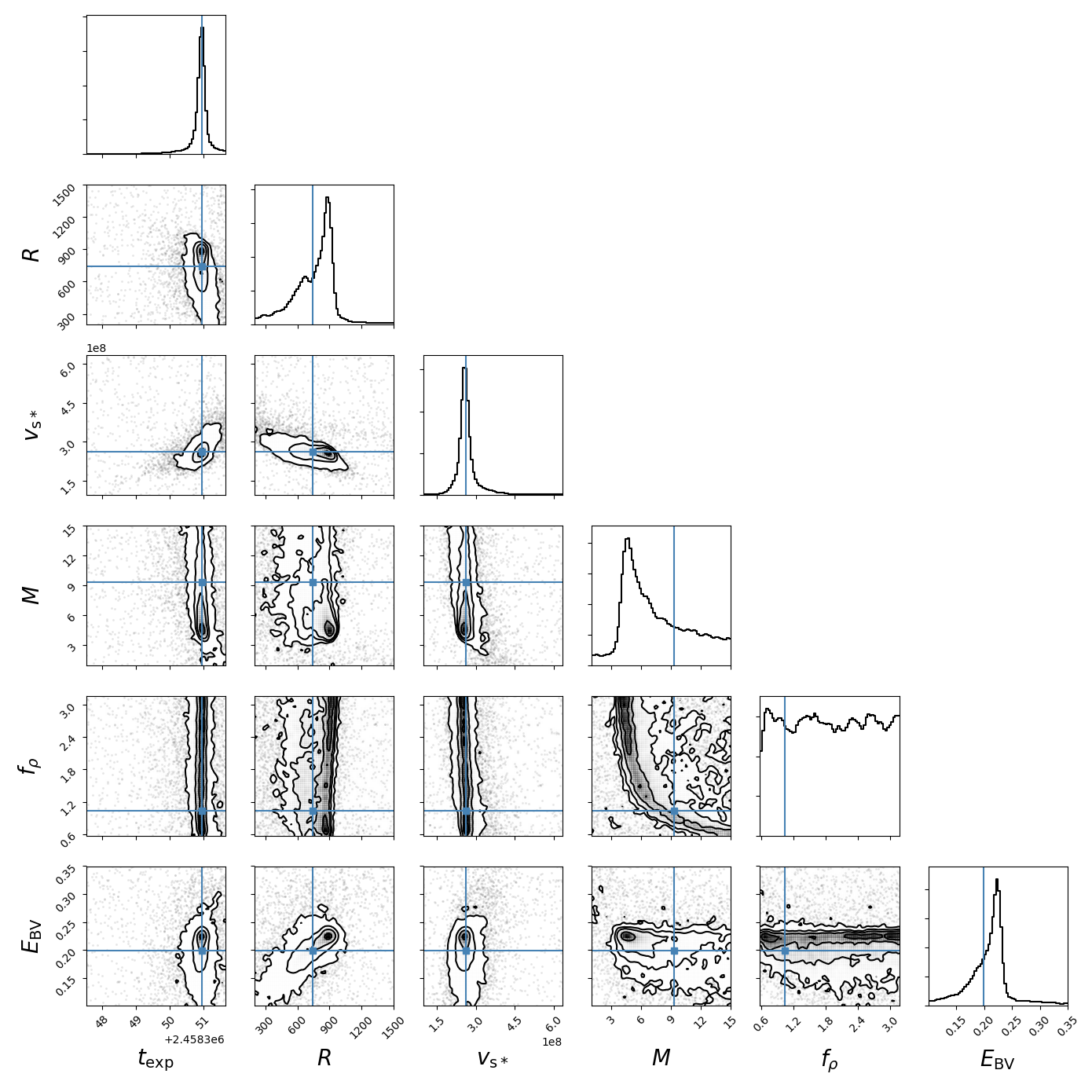}
\caption{One and two dimensional projections of the posterior probability distributions of the parameters $R$, $v_{\rm s*,8.5}$, $M$, $f_{\rm \rho}$, $t_{\rm ref}$, $E_{\rm B-V}$, demonstrating the covariance between parameters. The contours correspond to the 1$\sigma$, 2$\sigma$ and 3$\sigma$ symmetric percentiles. The blue line corresponds to the maximum of the posterior distribution, computed by the {\tt dynesty} nested sampling algorythm as part of the {\tt SOPRANOS-nested} package.}
\label{fig:triangle_constrained}
\end{figure*}

\begin{figure*}
\includegraphics[scale=.60]{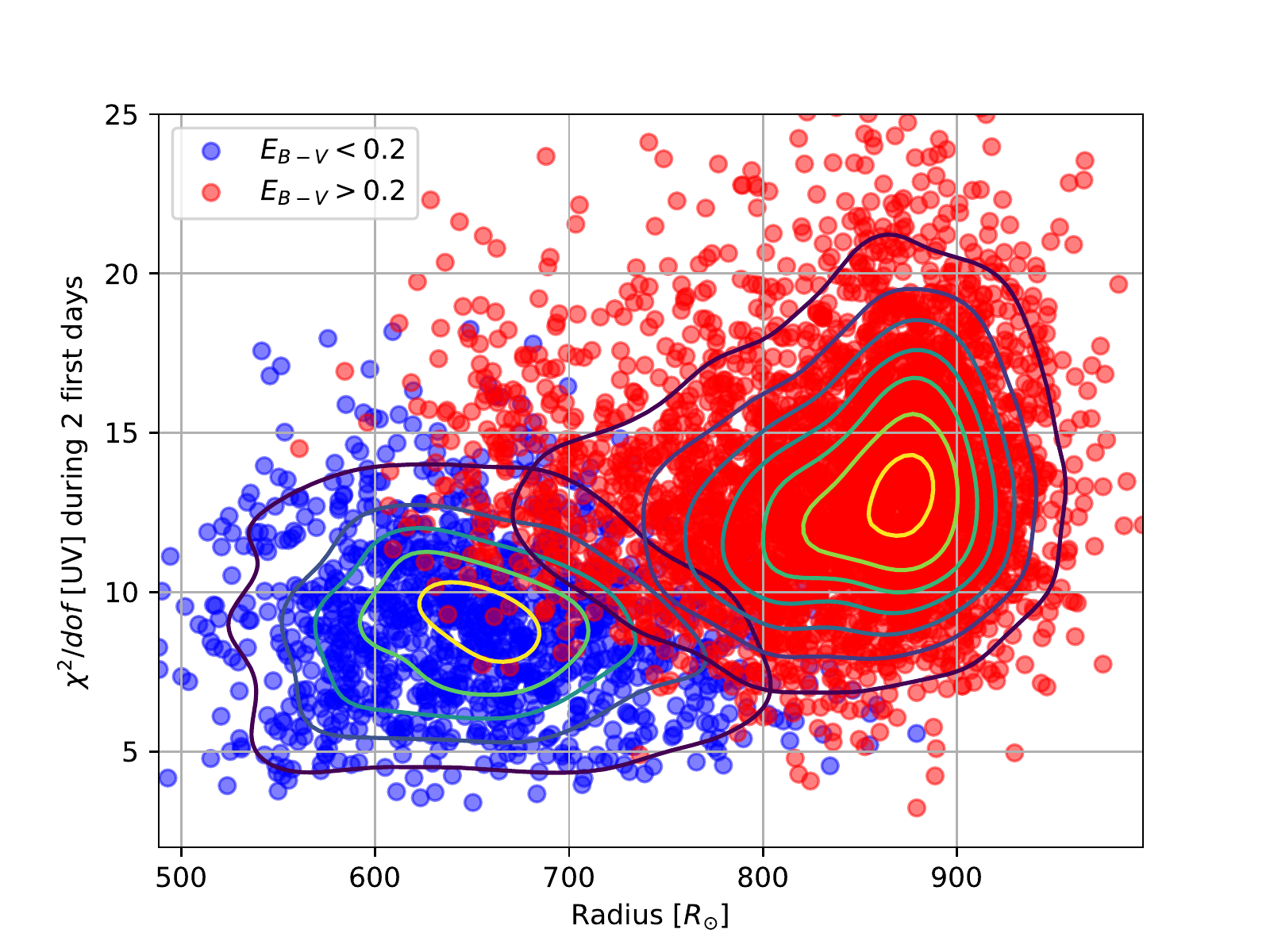}
\includegraphics[scale=.60]{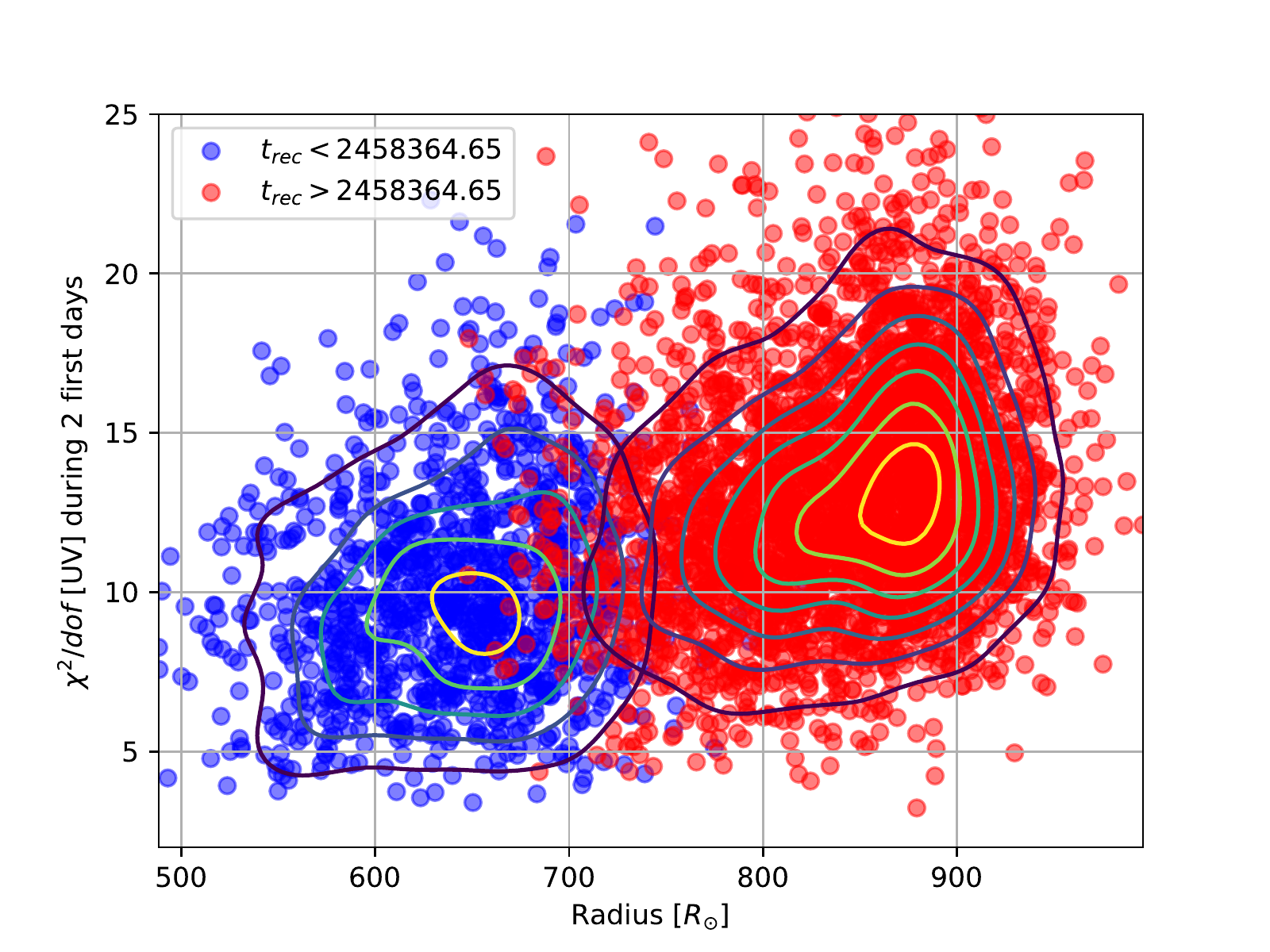}
\caption{$\chi^2/dof$ of all models in the {\tt dynesty} chain, considering only the first two days of UV data, as a function of the model progenitor's radius. The blue and red color coding distinguish between models with different values of (left panel) the extinction and (right panel) the time of recombination. The contours show the the density of models. Models (in blue) with $E_{\rm B-V}<0.2$ and $t_{\rm rec}<2458364.5$ are characterized by smaller radii and a better match the early UV data.}
\label{fig:underlying}
\end{figure*}


In Figure~\ref{fig:triangle_constrained}, we show the one and two dimensional projections of the $PDF$ distributions obtained by fitting our model to the data with {\tt SOPARANO-nested}.
A full tabulation of the best-fit parameters, as well as the $68.2\%$ confidence range for each parameter is shown in Table~\ref{table:results}. For the best-fit values, we report the maximum posterior distribution values computed by {\tt dynesty}
: $R=744.0_{-128.0}^{+183.0}\,R_{\odot}$, $M_{\rm ej}=9.3_{-5.8}^{+0.4}\,M_{\odot}$, $t_{\rm exp}=2458350.95_{-0.2}^{+0.13}\,\rm JD$, $f_{\rm \rho}=1.04_{-0.36}^{+1.4}$, $E_{\rm B-V}=0.199_{-0.019}^{+0.036}$, $v_{\rm s*,8.5}=0.828_{-0.118}^{+0.052}$, giving $\chi^2/dof=1.67$. In Figure~\ref{fig:lc_versus_model}, we show a comparison of the observed data and the best-fit model and in Figure~\ref{fig:evo_param_TR} we show a comparison of the blackbody temperature and radius measured from the data and predicted by the best-fit model.

Note that when the probability function is not purely Gaussian (e.g. if it is double-peaked, which is the case here) or is asymmetric, the maximum probability does not necessarily fall close to the median of the marginalized distributions. In particular, it can fall outside of the symmetric interval containing $68\%$ of the probability, which is often reported as the 1$\sigma$-confidence range, and does not reflect any asymmetry of the distribution. Here, we report instead the tightest intervals containing $68\%$ of the probability and including our best-fit values. We comment on the best-fit results below: 

\begin{deluxetable}{llll}
\tablecaption{Results of the model fitting}
\tablecolumns{4}
\tablewidth{0pt}
\tablehead{\colhead{Parameter}&\colhead{Best fit }&\colhead{Median}&\colhead{$68.2\%$ conf.}}
\startdata
\decimalcolnumbers
$R$&$744.29$&$804.8$&$[615.94,927.61]$\\
$v_{\rm s*,8.5}$&$0.828$&$0.817$&$[0.71,0.88]$\\
$M$&$9.3$&$6.7$&$[3.5,9.7]$\\
$t_{\rm exp}$&$2458350.95$&$2458350.95$&$[2458350.75,$\\
&&&$2458351.08]$\\
$E_{\rm B-V}$&$0.199$&$0.215$&$[0.18,0.235]$\\
$f_{\rm \rho}$&$1.04$&$1.86$&$[0.68,2.44]$\\
\hline
$\chi^2/dof$&$1.69$&$2.29$&$-$\\
&($263.24/156$)&($357.76/156$)&$-$\\
\enddata
\tablecomments{The table shows the best-fit parameters, the median values of the \st{MCMC} chains, and $68.2\%$ confidence range for each parameter, computed using the marginalised posterior distributions.}
\label{table:results}
\end{deluxetable}


\begin{itemize}
\item In Figure~\ref{fig:L_R_Davies}, we show red supergiant (RSG) radii and luminosities derived from the temperatures and luminosities measured by \cite{Davies2018} for RSGs in the small and large Magellanic Clouds (SMC and LMC). The best-fit value of the radius we find for the SN\,2018\,fif progenitor star, $R=744.0_{-128.0}^{+183.0}\,R_{\odot}$, is well within, but on the large side of the range of radii measured for RSGs. An important caveat to this comparison is that it holds if the host Galaxy of SN\,2018\,fif has a similar metallicity to the Magellanic clouds (since metallicity affects mass-loss and thereby mass, and radius). As can be seen in Figure~\ref{fig:host}, the SN is located at the outskirts of a spiral galaxy. We do not have sufficient data to estimate the metallicity at the explosion cite, but assuming that the galaxy is similar to the Milky Way and the usual metallicity gradient in spirals, we consider a sub-solar metallicity to be reasonable.

\item The value of $t_{\rm ref}$, the reference time of our model, is earlier than $t_0= 2458351.6537_{-0.0903}^{+0.0356}\,\rm JD$, the estimated epoch at which the extrapolated r-band light curve turns to zero. This is not surprising: $t_0$ is a measurement of the epoch of first-light in the {\it r-band} and hot young SNe are predicted to emit light in the UV before they significantly emit optical light: there is no reason for $t_0$ and $t_{\rm BO}$ to be strictly identical. 


\item The best-fit value of the extinction $E_{\rm B-V}=0.199_{-0.019}^{+0.036}\,mag$ is high: note that it is the sum of the galactic extinction $E_{\rm B-V}=0.10$ (deduced from \citealt{Schlafly2011} and using \citealt{Cardelli1989} extinction curves) and all other sources of extinction along the line of sight, including the extinction from the SN host galaxy. The galactic extinction has a relatively high contribution to the derived value of $E_{\rm B-V}$. Moreover, we used the effective wavelength of the NaD lines (in the Gemini Spectrum from August 21) in order to estimate the extinction from the host galaxy, following the relation by \cite{Poznanski2012}. We found that an estimate of the host extinction is $E_{\rm B-V,host}=0.10\pm 0.04$ which, summed with the galactic extinction, is consistent with the value of $E_{\rm B-V}$ we derived.

\item In order to verify whether our best-fit value for $v_{\rm s*,8.5}$ is consistent with the observations, we make an estimate of v$_{sh}$ using equations~\ref{eq:sw2} and equation 11, and 4 from \cite{Rabinak2011}, which provide an expression of the depth $\delta$ as a function of our model parameters and link it to $v_{\rm s*,8.5}$ and v$_{sh}$. \cite{Rabinak2011} also link the velocity of the shock to the velocity of the photosphere. We obtain that the predicted value of the velocity of the photosphere is between $10900$ and $12500\rm km\,s^{-1}$. Using the P-Cygni profile of the H line in the spectrum of SN\,2018\,fif at $t=+15.00\,\rm{days}$, we estimate the observed velocity $v\approx 10\,000\,{\rm km\,s^{-1}}$ and find that it is consistent with the model prediction.

\item A connection between $M$ and $R$ exists for hydrostatic stars undergoing secular evolution. This may not be the case for the progenitor just prior to explosion, in particular since it may have lost some mass just shortly prior to exploding. Here, we just report the constraint which the data impose on the parameters of this model, without assuming anything about the status of the progenitor.
\end{itemize}

Figure~\ref{fig:triangle_constrained} shows that the marginalized posterior distribution of radii is double-peaked. In Figure~\ref{fig:underlying}, we show the distribution of radii of all the models in the chain and the goodness-of-fit ($\chi^2/dof$) computed with the first two days of UV data, with a color code for different values of the extinction and recombination time. We find that models with lower progenitor radii appear to be characterized by lower values of the extinction ($E_{\rm B-V}<E_{\rm B-V,lim}=0.2$), earlier recombination time ($t_{\rm rec}<t_{\rm rec,lim}=2458364.65$) and better match to the early UV data. In the appendix, we show a full tabulation of the results of running {\tt SOPARANO-nested} with narrower priors on the extinction, $E_{\rm B-V} \in [0.15,0.2]$ and $E_{\rm B-V} \in [0.2,0.35]$, confirming that the models with lower values of $E_{\rm B-V}$ better match the UV data, and in particular the early UV data. 

Although the first spectrum showing clear observational signs that recombination has happened (H lines with strong P-cygni profiles) was taken at $t=2458366.65$ (two days after $t_{\rm rec,lim}$), it is reasonable to assume that recombination has happened several days before and that SN\,2018\,fif belongs to the class of objects with lower radii. If the correlations exhibited in Figure~\ref{fig:underlying} are confirmed with future objects, precisely constraining the time of recombination with denser spectroscopic measurement may help break the extinction/radius degeneracy. As the early UV data seem to distinguish between the two classes of objects, observations of early UV with higher accuracy (e.g. with ULTRASAT) may also enable one to remove the extinction/radius degeneracy, and independently determine both.

\subsection{Discussion}
Our modeling approach only uses the early part of the light curve. This makes sense, as we only aim at constraining a very limited set of progenitor parameters, mainly its radius $R$ and  $E/M_{\rm ej}$, which have been shown (by SW17, using numerical calculations) to determine the early light curve, independently of the stellar density profile or the explosion models uncertainties.

However, different modeling approaches exist. In particular, the use of numerical and radiation hydrodynamic codes can allow one to utilize the full light curve, until the late nebular phase, and can be very informative. Below, we give a few examples of recent works adopting or exploring this modeling approach, in order to put our own modeling choices within a broader context.

Several recent works have used radiation hydrodynamic codes to fit models of exploded progenitors to observed light curves. For example, \cite{Ricks2019} modeled the full light curve of eight supernovae discovered between 1999 and 2016. They used the stellar evolution code {\tt MESA} \citep{Paxton2011,Paxton2013,Paxton2015,Paxton2018,Paxton2019} and the radiative transfer code {\tt STELLA} \citep{Blinnikov1998,Blinnikov2004,Baklanov2005,Blinnikov2006} to evolve stars, explode them, and model their bolometric and individual-bands light curves. Assessing the goodness of the fit can be done by fitting the modeled light curves to those of observed SNe for which  pre-explosion imaging exist. This strategy was adopted e.g. by \cite{Bersten2019}, who modeled the light curve of six SNe which have direct progenitor detection, using a 1D Lagrangian hydrodynamical code. 
\cite{Eldridge2019} used the STAR code \citep{Eldridge2017} to model the light curves of eleven SNe with pre-explosion imaging. 
This approach contrasts with ours by the amount of parameters it aims at constraining. Indeed, the fitted parameters include the zero age main sequence mass of the progenitor, its rotation velocity, the initial metallicity, parameters governing the mass-loss of the star, its core mass, the wind properties that gave rise to the circumstellar material, the explosion energy, the amount of radioactive material synthesised in the explosion as well as its degree of mixing into the outer layers in the ejecta. 


Recently, hydrodynamical simulations have suggested that a degeneracy exists, beyond $10$ to $20$ days, between the light curves of different families of progenitors, underlying the constraining power of the earliest phase of the lightcurve. \cite{Goldberg2019} used {\tt MESA} and {\tt STELLA} to show how various families of progenitors produce light curves with similar observables, and explored whether this degeneracy could be broken. \cite{Dessart2019} used a 1D Lagrangian hydrodynamical code \citep{Livne1993,Dessart2010} to model stars of different mass in order to check whether the SN lightcurves they produce are different. They found that the different modeled progenitors produced similar lightcurves between 10 and 100 days and concluded that comparing models and light curves during this phase is not enough to deduce a unique model of the progenitor. This conclusion is in contrast with other works e.g. by \cite{Eldridge2019} -- who claim that it is possible to achieve strong constraints on the progenitors of type IIp supernovae from the light curves alone. Recently, \cite{Goldberg2020} showed that after the first $20$ days, families of explosion models with a wide range of $M_{\rm ej}$, $R$, and $E$ show a good match to the data of five SNe, and that pre-explosion imaging or modeling of the earlier shock cooling phase are key to properly constrain the progenitor's properties.

Our approach is therefore complimentary to the radiation hydrodynamic modeling approach. It aims at constraining a far smaller set of parameters, at stages of the explosion when radiation hydrodynamic codes often fail to properly model the light curve and before the degeneracy between the light curve of different progenitors becomes an obstacle to their modeling. Combining modeling of the shock-cooling phase with the radiation hydrodynamic modeling approach can break the degeneracy and allow one to use the assets of both approaches for a complete modeling of the progenitor properties.



\section{Conclusions}\label{sec:discussion}
We presented the UV and visible-light observations of SN~2018\,fif by ZTF and \textit{Swift}.  
The analysis of the early spectroscopic observations of SN\,2018\,fif reveals that its progenitor was surrounded by relatively small amounts of circumstellar material compared to a handful of previous cases. This particularity, as well as the high cadence multiple-bands coverage, make it  a good candidate to test shock-cooling models. 

We employed the {\tt SOPRANOS} code, an implementation of the model by \cite{SW2017} and its extension to early times by Morag, Sapir \& Waxman (M20; Morag et al. in preparation). The {\tt SOPRANOS} algorithm has the advantage of including a careful account for the limited temporal validity of the shock-cooling model (in this sense, our approach is similar to the one adopted by \citealt{Rubin2017}) as well as allowing usage of the the entirety of the early UV data through the M20 extension.

We find that - within the assumptions of the \cite{SW2017} model - the progenitor of SN 2018\,fif was a large red supergiant, with a radius of $R/R_{\odot}=744.0_{-128.0}^{+183.0}$ and an ejected mass of $M/M_{\odot}=9.3_{-5.8}^{+0.4}$. We find that the distribution of radii is double-peaked, with lower radii corresponding to lower values of the extinction, earlier recombination times, and better match to the early UV data. Our model also gives information on the explosion epoch, the progenitor inner structure, the shock velocity and the extinction.

Our approach aims at modeling a limited amount of key progenitor properties, mainly its radius $R$ and  $E/M_{\rm ej}$, using the constraining power of the early stages of the light curve. In this sense it is complementary to recent works that use numerical radiation hydrodynamic codes to model the later stages of the light curve and suffer from the degeneracy between the light curves of various progenitors at later stages (e.g. \citealt{Goldberg2020}).

As new wide-field transient surveys such as the Zwicky Transient Facility (e.g., \citealt{Bellm2019,Graham2019}) are deployed, many more SNe will be observed early, and quickly followed up with early spectroscopic observations and multiple-band photometric observations. The ULTRASAT UV satellite mission \citep{Sagiv2014} will collect early UV light curves of hundreds of core-collapse
supernovae. Their high accuracy may enable one to remove the extinction/radius degeneracy, and independently determine both. The methodology proposed in this paper offers a framework to analyze these objects, in order to constrain the properties of their massive progenitors and pave the way to a comprehensive understanding of the final evolution and explosive death of massive stars.

\software{{\tt ZTF pipeline} \citep{Masci2019},
          {\tt ZOGY} \citep{Zackay2016},
          {\tt HEAsoft} (v6.26, \citealt{Heasarc2014}),
          {\tt IRAF} \citep{Tody1986,Tody1993},
          {\tt dynesty} \citep{Skilling2004,Skilling2006,Higson2019,Speagle2020},
          {\tt LRIS pipeline} \citep{Perley2019},
          {\tt Astropy} \citep{Astropy2013,Astropy2018},
          {\tt Matplotlib} \citep{Hunter2007},
          {\tt Scipy} \citep{Virtanen2020}.
}

\acknowledgments

We dedicate this paper to the memory of Rona Ramon.

M.T.S. acknowledges support by a grant from IMOS/ISA, the Ilan Ramon fellowship from the Israel Ministry of Science and Technology and the Benoziyo center for Astrophysics at the Weizmann Institute of Science.

E.O.O is grateful for the support by grants from the Israel Science Foundation, Minerva, Israeli Ministry of Science, the US-Israel Binational Science Foundation, the Weizmann Institute and the I-CORE Program of the Planning and Budgeting Committee and the Israel Science Foundation.

AGY’s research is supported by the EU via ERC grant No. 725161, the ISF GW excellence center, an IMOS space infrastructure grant and the BSF Transformative program as well as The Benoziyo Endowment Fund for the Advancement of Science, the Deloro Institute for Advanced Research in Space and Optics, The Veronika A. Rabl Physics Discretionary Fund, Paul and Tina Gardner and the WIS-CIT joint research grant;  AGY is the recipient of the Helen and Martin Kimmel Award for Innovative Investigation.

The data presented here are based - in part - on observations obtained with the Samuel Oschin Telescope 48-inch and the 60-inch Telescope at the Palomar Observatory as part of the Zwicky Transient Facility project. ZTF is supported by the National Science Foundation under Grant No. AST-1440341 and a collaboration including Caltech, IPAC, the Weizmann Institute for Science, the Oskar Klein Center at Stockholm University, the University of Maryland, the University of Washington, Deutsches Elektronen-Synchrotron and Humboldt University, Los Alamos National Laboratories, the TANGO Consortium of Taiwan, the University of Wisconsin at Milwaukee, and Lawrence Berkeley National Laboratories. Operations are conducted by COO, IPAC, and UW.

We acknowledge the use of public data from the Swift data archive.

SED Machine is based upon work supported by the National Science Foundation under Grant No. 1106171

The data presented here were obtained - in part - with ALFOSC, which is provided by the Instituto de Astrofisica de Andalucia (IAA) under a joint agreement with the University of Copenhagen and NOTSA.The Liverpool Telescope, located on the island of La Palma in the Spanish Observatorio del Roque de los Muchachos of the Instituto de Astrofisica de Canarias, is operated by Liverpool John Moores University with financial support from the UK Science and Technology Facilities
Council.The ACAM spectroscopy was obtained as part of OPT/2018B/011.

\appendix
\section{Release of the {\tt PhotoFit} code}
The {\tt PhotoFit} tool, used to make Figure~\ref{fig:seds}, Figure~\ref{fig:evo_param_TR} and Figure~\ref{fig:evo_param_L} of this paper, is made available at {\tt https://github.com/maayane/PhotoFit}. {\tt PhotoFit} is a package for calculating and visualizing the evolution in time of the effective radius, temperature and luminosity of a supernova - or any target assumed to behave as a blackbody - from multiple-bands photometry. 

Measurements in different bands are usually taken at different epochs. The first task completed by {\tt PhotoFit} is to interpolate the flux and the errors on common epochs defined by the user. 

{\tt PhotoFit} then fits each SED with a blackbody model after (1) correcting for the extinction: PhotoFit does this using \cite{Schlafly2011} and using the extinction curves of \cite{Cardelli1989}; (2) correcting for the redshift (3) correcting for the effect of the filters transmission curves: PhotoFit does this using the {\tt pyphot} package\footnote{http://mfouesneau.github.io/docs/pyphot/} for synthetic photometry (Fouesneau, in preparation).

The fit itself can be done in two different ways (to be chosen by the user and defined in the params.py file):
\begin{itemize}
\item Nested sampling with {\tt dynesty} \citep{Skilling2004,Skilling2006,Higson2019,Speagle2020}.
\item A linear fit with a grid of temperatures. 
\end{itemize}

\section{Tabulation of the solution with narrow priors on $E_{B-V}$}

Here, we show a full tabulation of the results of running {\tt SOPRANOS-nested} applying narrow priors on the extinction: $E_{B-V}\in [0.15,0.20]$ and $E_{B-V}\in [0.20, 0.35]$. The lower extinction case gives a smaller best-fit radius, and a better match to the UV data, specifically the early (first two days) UV data.
\movetabledown=30mm
\begin{rotatetable}
\begin{deluxetable}{|l|lll|lll|}
\tablecaption{Results of the model fitting with narrow priors on $E_{B-V}$}
\tablecolumns{5}
\tablewidth{0pt}
\tablehead{
\colhead{}&\colhead{}&\colhead{$E_{B-V}\in [0.15,0.20]$}&\colhead{}&\colhead{}&\colhead{$E_{B-V}\in [0.20,0.35]$}\\
\colhead{Parameter}&
\colhead{Best fit }&
\colhead{Median}&\colhead{$68.2\%$ conf.}&\colhead{Best fit }&
\colhead{Median}&\colhead{$68.2\%$ conf.}}
\startdata
\decimalcolnumbers
$R$&$533.1$&$656.29$&$[486.4,720.12]$&$816.29$&$849.93$&$[746.16,979.86]$\\
$v_{\rm s*,8.5}$&$0.796$&$0.831$&$[0.745,0.88]$&$0.793$&$0.821$&$[0.71,0.879]$\\
$M$&$13.9$&$7.6$&$[5.8,14.2]$&$6.9$&$6.1$&$[3.2,9.4]$\\
$t_{\rm exp}$&$2458350.91$&$2458350.93$&$[2458350.82,$&$2458350.94$&$2458350.95$&$[2458350.74$\\
&&&$2458351.07]$&&&$,2458351.07]$\\
$E_{\rm B-V}$&$0.151$&$0.188$&$[0.15,0.194]$&$0.217$&$0.223$&$[0.2,0.23]$\\
$f_{\rm \rho}$&$1.24$&$1.81$&$[0.68,2.34]$&$1.88$&$1.8$&$[0.58,2.34]$\\
\hline
$\chi^2/dof$&$1.77$ ($264.45/149$)&$2.28$ ($339.60/149$)&$-$&$1.69$ ($263.99/156$)&$1.98$ ($332.95/168$)&$-$\\
$\chi^2/dof\,[UV]$&$1.70$ ($52.76/31$)&$1.51$ ($46.94/31$)&$-$&$2.40$ ($74.48/31$)&$1.78$ ($60.68/34$)&$-$\\
$\chi^2/dof$&&&$-$&&&$-$\\
$[UV,2\,days]$&$4.64$ ($9.28/2$)&$6.89$ ($13.77/2$)&$-$&$16.36$ ($32.72/2$)&$12.79$ ($25.57/2$)&$-$\\
\enddata
\tablecomments{The table shows the best-fit parameters, the median values of the dynesty chain, and $68.2\%$ confidence range for each parameter, computed using the marginalised posterior distributions. The two last lines of the table are (respectively) $\chi^2/dof$ using only the UV data and only the first two days of UV data.} %
\label{table:results_ext}
\end{deluxetable}
\end{rotatetable}
\bibliographystyle{aasjournal} 
\bibliography{bibliograph.bib}

\end{document}